\pdfoutput=1
\documentclass[11pt,aps,a4paper,eqsecnum,amsmath,amssymb,
  notitlepage,nofootinbib,
  longbibliography
  ]{revtex4-1}
\usepackage{graphicx}
\usepackage{tikz}

\def \bra #1{\langle #1 | }
\def \ket #1{ | #1 \rangle }
\def\Wop[#1][#2][#3][#4][#5]{W
  \left(\left.\begin{array}{cc}
        #1&#2\\#3&#4\end{array}\,\right|\,#5
  \right)}

\newcommand{\bolw}[5]{\left( \left.\begin{matrix}
                             #1 & #2 \\
                             #3 & #4 \\
                            \end{matrix}\,\right|\, #5 \right) }

\def \dim {\mathrm{dim}}

\begin{document}
\preprint{} 

\title{Exact solution of the $D_3$ non-Abelian anyon chain}

\author{Natalia Braylovskaya}
\author{Peter E. Finch}
\author{Holger Frahm}
\affiliation{%
Institut f\"ur Theoretische Physik, Leibniz Universit\"at Hannover,
Appelstra\ss{}e 2, 30167 Hannover, Germany}

\date{June 2, 2016}

\begin{abstract}
  Commuting transfer matrices for linear chains of interacting non-Abelian
  anyons from the two-dimensional irreducible representation of the dihedral
  group $D_3$ (or, equivalently, the integer sector of the $su(2)_4$ spin-$1$
  chain) are constructed using the spin-anyon correspondence to a
  $D_3$-symmetric formulation of the XXZ Heisenberg spin chain.  The spectral
  problem is solved using discrete inversion identities satisfied by these
  transfer matrices and functional Bethe ansatz methods.  The resulting
  spectrum can be related to that of the XXZ spin-$1/2$ Heisenberg chain with
  boundary conditions depending on the topological sector of the anyon chain.
  The properties of this model in the critical regime are studied by finite
  size analysis of the spectrum.  In particular, points in the phase diagram
  where the anyon chain realizes some of the rational $\mathbb{Z}_2$ orbifold
  theories are identified.
\end{abstract}

\pacs{%
05.30.Pr, 
05.70.Jk, 
03.65.Vf 
}

\maketitle

\section{Introduction}
There has been growing interest recently in the properties of the
exotic quasi-particles arising in topologically ordered systems such
as certain fractional quantum Hall states \cite{MoRe91} or
two-dimensional frustrated quantum magnets
\cite{MoSo01,BaFG02,Kita06}.  One-dimensional lattice models for these
non-Abelian anyons have been constructed starting from a consistent
set of fusion and braiding rules and the collective phases realized in
many-body anyon models with pairwise interactions have been studied,
mostly numerically, to identify a variety of critical phases and the
conformal field theories describing their low energy properties, see
e.g.\ \cite{FTLT07,TAFH08,GATH13,Finch.etal14}.  The gapless states in
these phases are protected by the topological symmetry present in
these chains and can be realized at interfaces between different
topological phases where they provide insights into nature of the
adjacent quantum liquids \cite{GATL09,GrSc09,BaSl09}.

Integrable models are known to be another important source
(complementary to numerical approaches) of un-biased information on
the properties of correlated many-body systems, in particular in
low-dimensional systems where quantum fluctuations are strong.  In
fact, some of the anyonic lattice models mentioned above can be
derived in the Hamiltonian limit of a class of integrable statistical
lattice models with interactions round the face (also called IRF or
face models), such as the restricted solid on solid (RSOS) models
\cite{AnBF84} and their generalizations \cite{Pasq88,Gepn92}.
Motivated by this connection several integrable chains of interacting
non-Abelian anyons have been constructed and studied recently, see
e.g.\ \cite{FiFr13,FiFF14}.

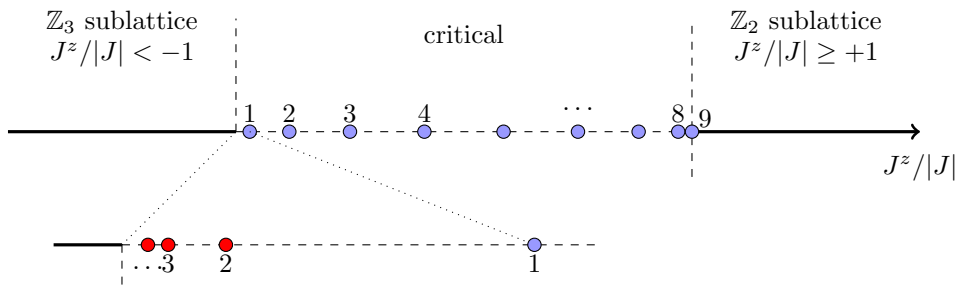
\begin{figure}[t]
  \begin{center}
    {\small
      \begin{tikzpicture}[scale=3.]
        \def \offs {.0}
        \def \offyp {-0.5}
        \def \offxp {28.5}
        \def \offsp {30}

        \tikzstyle{every node}=[circle,draw,thin,fill=blue!40,minimum
        size=5pt,inner sep=0pt] 
        \tikzstyle{every loop}=[]

        \coordinate [label=above:$\mathbb{Z}_3$ sublattice] (Z3) at
        (-1.5,0.15);
        \coordinate [label=above:$J^z/|J|<-1$] (Z3j) at (-1.5,0.02);
        \coordinate [label=above:$\mathbb{Z}_2$ sublattice] (Z2) at
        (1.5,0.15); 
        \coordinate [label=above:$J^z/|J|\ge+1$] (Z2j) at (1.5,0.02);
        \coordinate [label=above:critical] (Zc) at
        (0,0.25); 
        \coordinate [label=below:$J^z/|J|$] (Zc) at (2,0.02); 

        \draw[very thick]  (-2,0) -- (-1,0);
        \draw[dashed]  (-1,0) -- ( 1,0);
        \draw[very thick,->]  (1,0) -- ( 2,0);
        \draw[dashed] (-1.0,-0.0) -- (-1.0,0.5);
        \draw[dashed] (1.0,-0.2) -- (1.0,0.5);
        \node [label=above:$1$] at (-0.939693,\offs) {};
        \node [label=above:$2$] at (-0.766044,\offs) {};
        \node [label=above:$3$] at (-0.5     ,\offs) {};
        \node [label=above:$4$] at (-0.173648,\offs) {};
        \node at (0.173648,\offs) {};
        \node [label=above:$\ldots$] at (0.5     ,\offs) {};
        \node at (0.766044,\offs) {};
        \node [label=above:$8$] at (0.939693,\offs) {};
        \node [label=above right:$9$] at (1,0.0) {};
        %

\draw[very thick] (\offxp-1*\offsp,\offyp) -- (\offxp-1.01*\offsp,\offyp);
\draw[dashed]  (\offxp-1*\offsp,\offyp) -- (\offxp-0.93*\offsp,\offyp);
\draw[dotted]  (-0.93969,0.) -- (\offxp-0.93969*\offsp,\offyp);
\draw[dotted]  (-1.,0.) -- (\offxp-1.*\offsp,\offyp);
\draw[dashed]  (\offxp-1.*\offsp,\offyp) -- +(0.,-0.2);
\node [label=below:$1$] at (\offxp-0.9396926208*\offsp,\offyp) {};
\node [fill=red,label=below:$2$] at (\offxp-0.9848077530*\offsp,\offyp) {};
\node [fill=red,label=below:$3$] at (\offxp-0.9932383577*\offsp,\offyp) {};
\node [fill=red,label=below:$\ldots$] at (\offxp-0.9961946981*\offsp,\offyp) {};
      \end{tikzpicture}
    }
  \end{center}
  \caption{Phase diagram of the $D_3$ anyon model: blue circles
    indicate the location of the points in the critical region
    corresponding to the rational $\mathbb{Z}_2$ orbifold CFTs with
    compactification radii $\tilde{r}=\sqrt{p/2}$ for
    $p=1,2,3,\ldots,9$ as denoted by the labels.
    Similarly, red circles in the blow-up of the region around $J^z=-|J|$
    indicate the sequence of rational orbifold theories at the dual radii
    $\tilde{r}=\sqrt{1/2p}$ for $p=2,3,4,\ldots$\,.  
    The CFT realized at the self-dual point, 
    $p=1$, is the Kosterlitz-Thouless theory. \label{fig:d3-phases}}
\end{figure}

Here we consider one of the deformations of $su(2)$ quantum spin chains
leading to an anyonic models studied by Gils \emph{et al}.\ \cite{GATH13},
namely the $su(2)_{k=4}$ spin-$1$ or, equivalently, $D_3$ anyon model, see
also Refs.~\cite{MaPV10,VeMP11}.  
As pointed out in Ref.~\cite{Finch13} this model can be related to one of the
paradigms of integrable quantum chains, i.e.\ the XXZ Heisenberg spin chain,
under a particular spin-anyon (or face-vertex) correspondence
\cite{Pasq88,Roch90}.  Based on this correspondence we have been able to embed
the anyonic Hamiltonian into a family of commuting operators generated by an
IRF transfer matrix.  Interestingly, and unlike the usual $su(2)$ spin-$1$
chain with bilinear and biquadratic exchange or, e.g., the model of
interacting $so(5)_2$ non-Abelian anyons \cite{Finch.etal14}, this embedding
is possible for arbitrary coupling constants.  Specifically, the spectrum of
the anyon chain can be related to that of the XXZ spin chain subject to
properly chosen boundary conditions depending on the topological sector of the
former.
The independence of thermodynamic properties on these boundary conditions this
allows to relate the phase diagram of the anyon chain to that known from the
exact solution of the Heisenberg model, see Fig.~\ref{fig:d3-phases}.

We find that, while the low energy properties of the anyon and the
spin chain in the critical phases are both described by conformal
field theories (CFTs) with central charge $c=1$ their operator content
is different as a consequence of the change in boundary conditions and
the presence of additional selection rules on a $U(1)$ charge (i.e.\
the magnetization for the spin chain) for the anyon model: the
continuum limit of the critical Heisenberg model is known to a free
boson with compactification radius $r_G$ depending on the anisotropy.
The critical properties of the anyon model, on the other side are
$\mathbb{Z}_2$ orbifolds of a boson compactified on a circle with
radius $\tilde{r}=3\,r_G$.\footnote{%
  A similar relation exists for another model related to the XXZ
  chain, i.e.\ the critical Ashkin-Teller quantum chain \cite{KoNK81}:
  in the continuum limit its partition function is identical to that
  of the orbifold model with $\tilde{r}_{\mathrm{AT}}=2\,r_G$
  \cite{Yang87,Saleur_AT87}.}
For rational values of $\tilde{r}^2$ these CFTs have an extended symmetry
allowing for their formulation in terms of a finite number of primary fields.
From the exact solution of the $D_3$ anyon chain the location of these special
points in the phase diagram can be given in terms of the coupling constant of
the microscopic model.  These include the special rational CFTs corresponding
to compactification radii $\tilde{r}=\sqrt{p/2}$ with integer $p=1,2,\ldots,9$
that have been identified previously in the numerical investigation of
Ref.~\cite{GATH13}, see Fig.~\ref{fig:d3-phases}.  We note that the sequence
of these field theories for \emph{arbitrary} integer $p$ is realized for the
coupling constant corresponding to the dual radius, $\tilde{r} \leftrightarrow
1/(2\tilde{r})$.

The paper is organized as follows: first we recall some basic facts related to
the integrability of the XXZ Heisenberg chain (or six-vertex model) and its
formulation as a $D_3$ symmetric spin chain.  The $D_3$ group algebra can be
extended to form a quasi-triangular Hopf algebra which allows to construct
related chains of anyons obeying $D_3$ fusion rules with an integrability
structure closely related to that of the XXZ model.  Based on this structure
we employ functional Bethe ansatz methods to solve the spectral problem of the
$D_3$ anyon model.  Using this solution we perfom a finite-size analysis of
the spectrum in the critical regime to identify the field theory describing
the collective behaviour of the anyons at low energies.

\section{The Heisenberg Spin-$1/2$ Chain}

The XXZ Heisenberg Hamiltonian for spin-1/2 particles with nearest neighbor
interaction on a one dimensional lattice and for periodic boundary conditions
is given as
\begin{equation}
  \label{ham-xxz}
 H=\frac{1}{2}\sum\limits_{i}\left(
   J(\sigma^{x}_{i}\sigma^{x}_{i+1}+\sigma^{y}_{i}\sigma^{y}_{i+1})
   +J^{z}\sigma^{z}_{i}\sigma^{z}_{i+1} 
  \right)\,. 
\end{equation}
This model is closely connected to the classical integrable six-vertex model
on the square lattice.  It is defined in terms of local Boltzmann weights
defined for each vertex with the state variables on the links connected to it.
The integrability of this model relies on the Yang-Baxter-Equation (YBE)
satisfied by the vertex weights,
\begin{equation}
  R_{12}(u-v)R_{13}(u)R_{23}(v)=R_{23}(v) R_{13}(u)R_{12}(u-v)
\end{equation}
For the 6-vertex-model the $R$-matrix is
\begin{equation}
  \label{rmat-6v}
  R(u)=\begin{pmatrix}
  a(u) &&& \\
  & b(u) & c(u) & \\
  & c(u) & b(u) & \\
  &&&a(u)\\
     \end{pmatrix}\,,
\end{equation}
where $a(u)=\sinh(u+i\gamma)$, $b(u)=\sinh(u)$, and $c(u)=\sinh(i\gamma)$.
Here $u$ is the spectral parameter, $\gamma$ parametrizes the anisotropy.  For
$u=0$ the $R$-matrix becomes proportional to a permutation-operator of two
sites.
As a consequence of the YBE the transfer matrix
\begin{equation}
  \tau(u)= tr_{0} \left(R_{01}(u)R_{02}(u)\cdots R_{0L}(u)\right)
\end{equation}
commutes for different arguments and therfore generates a family of commuting
operators.  Among these $\tau(0)\propto e^{i P}$ is proportional to the
translation operator allowing to define the momentum operator in the lattice
model, for periodic boundary conditions the eigenvalues of $P$ are
\begin{equation}
  P=\frac{2\pi}{L}n,\quad n\in \mathbb{N}\,.
\end{equation}
Higher integrals of motion are obtained by expansion of $\log\tau(u)$ around
$u=0$ with the XXZ Hamiltonian with $J^z/|J|=\cos\gamma$ given as
\begin{equation}
  H\propto \frac{\partial}{\partial u}\log\tau(u)|_{u=0} \,.
\end{equation}
Note that for even $L$ the XXZ Hamiltonian $H(J,J^z)$ is unitary equivalent to
$H(-J,J^z)$.  Therefore we will consider only the case $J=+1$ in most of the
discussion below.  Furthermore, the XXZ chain has a $U(1)$ symmetry resulting
in the conservation of the total magnetization $M=L/2-S^{z}$, where $S^{z}$ is
the $z$-component of the total spin.  Hence the eigenstates of the transfer
matrix can be assigned to different $M$-sectors.

\subsection*{The {XXZ} model as a $D_{3}$ spin chain}
Consider the dihedral group $D_{3}$, the group of symmetries of a triangle
generated by rotation $\sigma$ and reflection $\tau$: $D_{3}=\{\sigma,\tau
|\sigma^{3}=\tau^{2}=1\}$ \cite{Finch13}. This group has two one-dimensional
irreducible represnetations (irreps) $(\pi_{\pm},V_{\pm})$ and one
two-dimensional irrep $(\pi_{2},V_{2})$.
To formulate the XXZ model as a $D_3$-chain we identify the states spanning
the representation spaces $V_\alpha$ with the states $\ket{\sigma_1,\sigma_2}
\in\mathbb{C}^2\otimes\mathbb{C}^2$ of a system of two $SU(2)$ spins $\frac12$
as
\begin{equation*}
  V_{\pm}\propto\left\{\ket{\uparrow\downarrow}
                      \pm\ket{\downarrow\uparrow}\right\}\,,
  \quad
  V_{2}\propto\left\{\ket{\uparrow\uparrow},
                    \ket{\downarrow\downarrow}\right\}\,.
  \quad
\end{equation*}
Now both the nearest neighbour XXZ Hamiltonian (\ref{ham-xxz}) and the
$R$-matrix (\ref{rmat-6v}) of the six-vertex model can be represented in terms
of the projection operators $P^{(2,\pm)}$ onto the corresponding two-spin
states:
\begin{equation}
  \label{6v-proj}
  \begin{aligned}
    h_{i,i+1}&=(J^{z}-J)P_{i,i+1}^{(2)}-2J P_{i,i+1}^{(-)} \\
    R(u)&=
    \left(\mathrm{sinh}(u)+\mathrm{sinh}(i\gamma)\right)P^{(+)} 
    +\left(\mathrm{sinh}(u)-\mathrm{sinh}(i\gamma)\right)P^{(-)}
    + \mathrm{sinh}(u+i\gamma)P^{(2)}\,.
  \end{aligned}
\end{equation}
Note that the periodic closure is compatible with this construction, therefore
the XXZ Hamiltonian is $D_3$-symmetric.

\section{The integrable $D_3$ anyon chain}
\label{sec:intmodel}
\subsection{Anyon chain from braided fusion categories}
Mathematically, anyonic theories can be represented by braided tensor
categories.  An anyon model consists of a collection of conserved topological
objects $\psi_a$ obeying a commutative fusion algebra
\begin{equation}
  \psi_a\otimes\psi_b \simeq \oplus_c N_{ab}^c\, \psi_c
\end{equation}
with non-negative integers $N_{ab}^c$ (we assume that the tensor category is
multiplicity free which corresponds to $N_{ab}^c\in\{0,1\}$).  Fusion can be
represented graphically
\begin{center}
\begin{tikzpicture}
   \draw (0.4,1) -- (1,0.5) -- (1.6,1);
   \draw (1,0) -- (1,0.5);
   \node [above] at (0.3,1) {$\psi_a$};
   \node [above] at (1.7,1) {$\psi_b$};
   \node [below] at (1,0) {$\psi_{c}$};
\end{tikzpicture}
\end{center}
implying that $\psi_c$ appears in the fusion of $\psi_a$ and $\psi_b$,
i.e.\ $N_{ab}^c=1$.
Associativity of the algebra allows for a reordering of the fusion of multiple
anyons governed by so-called $F$-moves (the analog of $6$-$j$ symbols)
\begin{equation}
\label{fmoves}
\begin{tikzpicture}
   \draw (0+3,0) -- (3+3,0);
   \draw (1+3,0) -- (1+3,0.5);
   \draw (2+3,0) -- (2+3,0.5);
   \node [above] at (1+3,0.5) {$\psi_b$};
   \node [above] at (2+3,0.5) {$\psi_c$};
   \node [below] at (0.5+3,0) {$\psi_{a}$};
   \node [below] at (1.5+3,0) {$\psi_{d}$};
   \node [below] at (2.5+3,0) {$\psi_{e}$};
   %
   \node  at (8,0) {$=\sum\limits_{d^{\prime}}
     \left( F^{abc}_{e}\right)_{d'}^{d}$};
   \draw (0.2+10,0) -- (-0.2+10+3,0);
   \draw (1.5+10,0) -- (1.5+10,0.6);
   \draw (1.5+10,0.6) --(1.5+10-0.7,0.6+0.25);
   \draw (1.5+10,0.6) --(1.5+10+0.7,0.6+0.25);
   \node [above] at (1.5+10-0.9,0.3+0.25) {$\psi_b$};
   \node [above] at (1.5+10+0.9,0.3+0.25) {$\psi_c$};
   \node [below] at (0.5+10,0) {$\psi_a$};
   \node [right] at (1.6+10,0.4) {$\psi_{d'}$};
   \node [below] at (2.5+10,0) {$\psi_e$};
  
\end{tikzpicture}
\end{equation}

Here, we consider the fusion algebra obeyed by the finite-dimensional irreps
(or anyons) of $D_3$ \cite{Finch13}.  With $(\pi_+,V_{+})$ being the unique
'vacuum' the remaining non-trivial fusion rules are
\begin{equation}
V_{2}\otimes V_{2}=V_{+}\oplus V_{2}\oplus V_{-} \,,
\quad V_{-}\otimes V_{-}=V_{+}\,,
\quad V_{-}\otimes V_{2}=V_{2}\,.
\end{equation}
Below we shall identify representations spaces with the anyonic charges
labelled $2$, $\pm$, respectively.

The $D_3$ anyon model is constructed from $L$ copies of the two-dimensional
anyon with charge $2$ (which is, in fact, identical to the integer sector of
the $su(2)_4$ spin-$1$ anyon chain considered in
Refs.~\cite{MaPV10,VeMP11,GATH13}). 

\begin{figure}[t]
  \begin{center}
    {\small
      \begin{tikzpicture}[scale=0.7]
        \tikzstyle{every node}=[circle,draw,thin,fill=blue!40,minimum size=10pt,inner sep=0pt]
        \tikzstyle{every loop}=[]
        \node (n0) at (0.0,0.0) {$+$};
        \node (n1) at (3.0,0.0) {$2$};
        \node (n2) at (6,0.0) {${-}$};
        \foreach \from/\to in {n0/n1,n1/n2} \draw (\from) -- (\to);
        \draw (n1) .. controls (0,3) and (6,3) .. (n1);
      \end{tikzpicture}
    }
  \end{center}
\caption{Graphical representation of allowed neighbouring anyon labels in
  fusion paths of $D_3$ anyons of type $2=(\pi_2,V_2)$. Nodes of the graph
  correspond to the possible types of anyons which are connected via an edge
  if the two anyon labels can appear next to each other. \label{fig:d3-adj}}
\end{figure}
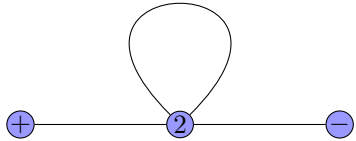

Based on the fusion rules we can construct the fusion path basis of the
anyonic Hilbert space: the states $\ket{a} \equiv |a_0 a_1 a_2 \ldots
a_L\rangle$ correspond to a sequence of anyonic charges subject to the
constraint $a_{i+1}$ appears in $a_i\otimes 2$.  It is convenient to represent
these graphically as
\begin{center}
\begin{tikzpicture}
 
\draw (-0.5+16,0)--(4+16,0);
 \draw (0+16,0)--(0+16,0.5);
 \draw (1+16,0)--(1+16,0.5);
 \draw (2+16,0)--(2+16,0.5);
 \draw (3+16,0)--(3+16,0.5);
 
 \node [above] at (0+16,0.5) {$2$}; 
 \node [above] at (1+16,0.5) {$2$};
 \node [above] at (2+16,0.5) {$2$};
 \node [above] at (3+16,0.5) {$2$};

 \node [below] at (-0.5+16,0) {$a_{0}$};
 \node [below] at (0.5+16,0) {$a_{1}$};
 \node [below] at (1.5+16,0) {$a_{2}$};
 \node [below] at (2.5+16,0) {$a_{3}$};
 \node [below] at (3.5+16,0) {$a_{4}$};
\end{tikzpicture}
\end{center}
The vertices in this diagram represent the fusion of two incoming anyons (top
and left) resulting in the outgoing anyon (right).  The fusion rules lead to
a local condition on the possible neighbouring labels
\begin{equation}
  a_{i}a_{i+1}\in \{+2,-2,2+,2-,22\}
\end{equation}
which is conveniently presented in the form of a graph with an adjacency
matrix $A_{ab}=N_{a2}^{b}$, i.e.\
\begin{equation}
  \label{d3-adjmat}
  A = \left( \begin{array}{ccc}
      0&~1~&0\\
      1&1&1\\
      0&1&0
    \end{array}
  \right)
\end{equation}
for the present case, such that allowed pairs of labels correspond to
adjacent nodes, see Fig.~\ref{fig:d3-adj}.
This constraint together with periodic boundary conditions $a_0=a_L$ yields
the total dimension of the Hilbert space of the $L$-site anyon chain
$\dim(\mathcal{H}) = 2^{L}+(-1)^{L}$.

Nearest neighbour $D_3$-symmetric interactions can be expressed in terms of
two-site projection operators on the different topological charges.  They can
be expressed in terms of the $F$-moves (\ref{fmoves}) and their inverses
$\bar{F}$ as
\begin{equation}
  \tilde{P}^{(b)}_{i-1,i,i+1} =
  \left(\bar{F}_{a_{i+1}}^{a_{i-1}22}\right)_{a_i'}^b
  \left({F}_{a_{i+1}}^{a_{i-1}22}\right)_{b}^{a_i}
  \ket{a_{i-1}a_i'a_{i+1}}\bra{a_{i-1}a_ia_{i+1}}\,,\qquad b\in\{2,\pm\}.
\end{equation}
Note that an operator coupling anyons on neighbouring sites $i$ and $i+1$ acts
on three neighbouring labels, i.e.\
$O_{i,i+1}=\tilde{O}_{a_{i-1}a_{i}a_{i+1}}$ in the fusion path basis.  Only
the anyon label on site $i$ may change under its action though.

\subsection{Spin-anyon correspondence}
Based on these projection operators we can construct an anyonic quantum chain
by exploiting the correspondence in terms of the $D_3$ description
\cite{Finch13}.  As in (\ref{6v-proj}) we can define an $R$-matrix
\begin{equation}
  \begin{aligned}
    \tilde{R}_{i-1,i,i+1}(u)&= 
    \left(\sinh(u)+\sinh(i\gamma)\right)\tilde{P}^{(+)}_{i-1,i,i+1}
    +\left(\sinh(u)-\sinh(i\gamma)\right)\tilde{P}_{i-1,i,i+1}^{(-)}\,\\
    &\quad+\sinh(u+i\gamma)\tilde{P}_{i-1,i,i+1}^{(2)}\,.
  \end{aligned}
\end{equation}
The matrix elements of $\tilde{R}(u)$ are the Boltzmann weights $W$ of a
two-dimensional statistical model with interactions round the face (IRF) 
\begin{equation}
    \begin{tikzpicture}[scale=0.5]
      \node at (-7,11.3) {$\bra{abd} \tilde{R}(u) \ket{acd} \equiv
        W\bolw{a}{b}{c}{d}{u} =$};
      \draw (0,10)  --   +(2.5,0);
      \draw (2.5,10)-- +(0,2.5);
      \draw (2.5,12.5)-- +(-2.5,0);
      \draw (0,12.5)--   +(0,-2.5);
      \node [left] at (0,10) {{$c$}};
      \node [left] at (0,12.5) {{$a$}};
      \node [right] at (2.5,12.5) {{$b$}};
      \node [right] at (2.5,10) {{$d$}};
      \node at (1.3,11.3) {$u$};
    \end{tikzpicture}
\end{equation}
The non-zero ones are
\begin{equation}
  \label{bweights}
  \begin{split}
    & W\bolw{2}{2}{2}{2}{u}=c(u)\\
    & W\bolw{\pm}{2}{2}{2}{u}=W\bolw{2}{2}{2}{\pm}{u}=a(u)\\
    & W\bolw{2}{\pm}{2}{2}{u}=W\bolw{2}{2}{\pm}{2}{u}=\frac{1}{\sqrt{2}}b(u)\\
    & W\bolw{\pm}{2}{2}{\pm}{u}=\left( c(u)+b(u)\right)\,,\qquad
    \phantom{\frac12}
    W\bolw{\pm}{2}{2}{\mp}{u}=\left(c(u)-b(u)\right)\\
    & W\bolw{2}{\pm}{\pm}{2}{u}=\frac{1}{2}\left(c(u)+a(u)\right)\,,\qquad
    W\bolw{2}{\pm}{\mp}{2}{u}=\frac{1}{2}\left(c(u)-a(u)\right)\\
  \end{split}
\end{equation}
They satisfy the (face) Yang-Baxter equation
\begin{equation}
  \label{ybe-irf}
\begin{aligned}
  &\sum_{g}
  W\bolw{f} {a}{e}{g}{u-v} 
  W\bolw{a} {b}{g}{c}{v} W\bolw{g} {c}{e}{d}{u}\,\\
&\quad =\sum _{g^{\prime}}
  W\bolw{a} {b}{f}{g^{\prime}}{u} W\bolw{f} {g^{\prime}}{e}{d}{v}   
  W\bolw{b} {c}{g^{\prime}}{d}{u-v}\,.
  \end{aligned}
\end{equation}
Additional properties are of the Boltzmann weights are unitarity
\begin{equation}
  \label{id:unitarity}
  \sum_e W\bolw{a}{e}{d}{c}{u} W\bolw{a}{b}{e}{c}{-u} =
  \rho(u)\rho(-u)\,\delta_{bd}\, 
\end{equation}
with $\rho(u) = \sinh(u+i\gamma)$, and a generalized crossing symmetry

\scalebox{1.0}{ 
\begin{tikzpicture}
\def \rcs {1.5} 
\def \dcs {2.3}
\node [right] at (-4.2,0.5*\rcs) {};

\draw (0,0)--(0,\rcs)--(\rcs,\rcs)--(\rcs,0)--(0,0);

\draw (\dcs*\rcs,0)--(\dcs*\rcs,\rcs)--(\rcs+\dcs*\rcs,\rcs)--(\rcs+\dcs*\rcs,0)--(\dcs*\rcs,0);

\node [below left]  at (0,0) {$c$};
\node [above left]  at (0,\rcs)  {$a$};
\node [above right]  at (\rcs,\rcs)  {$b$};
\node [below right ]  at (\rcs,0) {$d$};

\node [below left] at (\dcs*\rcs,0) {$\bar{a}$};
\node [above left] at (\dcs*\rcs,\rcs) {$b$};
\node [above right] at (\rcs+\dcs*\rcs,\rcs) {$d$};
\node [below right] at (\rcs+\dcs*\rcs,0) {$\bar{c}$};

\node at (0.5*\rcs,0.5*\rcs) {$u$};
\node at (0.5*\rcs+\dcs*\rcs,0.5*\rcs) {$-u - i\gamma$};

\node [left] at (2.1*\rcs,0.5*\rcs) {$=\mathcal{G}_{[ab]}$};
\node [right] at (3.5*\rcs,0.5*\rcs) {$\left(\mathcal{G}_{[ca]}\right)^{-1}$};
\end{tikzpicture}}

\begin{equation}
  \label{id:crossing}
  W\bolw{a}{b}{c}{d}{u} = \mathcal{G}_{[ab]} \,
                        W\bolw{b}{d}{\bar{a}}{\bar{c}}{-u-i\gamma} \,
                        \left(\mathcal{G}_{[cd]} \right)^{-1}
\end{equation}
where we have defined a gauge transformation $\mathcal{G}$
\begin{equation}
  \mathcal{G}_{[\pm2]} = -\sqrt{2}\,,\quad
  \mathcal{G}_{[2\pm]} = -\frac{1}{\sqrt{2}}\,,\quad
  \mathcal{G}_{[22]} = 1\,,
\end{equation}
and the mapping $a\to \bar{a}\equiv f(a)$ with $f({\pm}) = \mp$, $f(2)=2$.

The initial conditions satisfied by the Boltzmann weights at the special
points $u=0,-i\gamma$ are
\begin{equation}
  \label{id:initval}
  W\bolw{a}{b}{c}{d}{0} = \sinh(i\gamma)\,\,\delta_{cb}\,, \qquad
  W\bolw{a}{b}{c}{d}{-i\gamma} = \left(\mathcal{G}_{[ca]}\right)^{-1}
    \mathcal{G}_{[db]}\,\sinh(i\gamma)\,\,\delta_{a\bar{d}}\,.
\end{equation}

As a consequence of the Yang-Baxter equation (\ref{ybe-irf}) the transfer
matrix (for periodic boundary conditions $a_{L+1}=a_L$, $b_{L+1}=b_L$)
\begin{equation}
  \tilde\tau(u) = W\bolw{a_1}{a_{2}}{b_1}{b_{2}}{u}
                  W\bolw{a_2}{a_{3}}{b_2}{b_{3}}{u}
                  \cdots\
                  W\bolw{a_L}{a_{1}}{b_L}{b_{1}}{u}
\end{equation}
commutes for different arguments and therefore generates a family of commuting
operators on the Hilbert space of the $D_3$ anyon chain.  The initial
conditions (\ref{id:initval}) imply that among these the momentum
operator is obtained from the generator of translations as
\begin{equation}
  \label{irf:momentum}
  p = -i\log\left(\tilde\tau(0) / \sinh^L(i\gamma) \right)\,.
\end{equation}
The Hamiltonian with local interactions is the logarithmic derivative of the
transfer matrix at $u=0$.  In terms of the $D_3$ projection operators in the
anyon basis it reads (up to a constant)
\begin{equation}
  \label{irf:hamil}
  \tilde{H} = \sinh(i\gamma)\,\partial_u\left.\log\tilde\tau(u)\right|_{u=0} = 
  \sum_{i} \left((\cos\gamma-1) \tilde{P}^{(2)}_{i-1,i,i+1}
                -2 \tilde{P}^{(-)}_{i-1,i,i+1}\right)\,,
\end{equation}
where we have set $J=+1$ and $J^z=\cos\gamma$.  The case $J=-1$ will be
discussed below in the context of the low energy effective theory.

\subsection{Symmetries}
From the Boltzmann weights (\ref{bweights}) it follows immediately that the
face model has a $Z_2$-symmetry generated by the mapping $a\to \bar{a}$ on
each site
\begin{equation}
  \label{z2par}
  \hat{\sigma} | a_1 a_2 \cdots a_L \rangle
  = | \bar{a}_1 \bar{a}_2 \cdots \bar{a}_L \rangle
\end{equation}
(for even $L$ there are, in fact, two such symmetries related to the
restriction of these parity operations on either of the two sublattices).

In addition, the spectrum of the anyon chain can be decomposed into sectors
with different topological charge $Y$ (related to the flux through a ring of
anyons \cite{GATL09,GATH13})
\begin{equation}
  \bra{x^{\prime}} Y
  \ket{x}=\prod\limits_{i=1}^{L}
  \left( F^{2 x_{i+1} 2}_{x_{i} ^{\prime}} \right) ^{x_{i+1}^{\prime}}_{x_{i}}
\end{equation} 
where $x_i\in \{\pm,2\}$ and $\ket{x}$ ($\bra{x}$) is a (dual) vector from the
anyonic fusion path basis.  
The $F$-moves can be related to the Boltzmann weights of the isotropic
($\gamma=0$) anyon model through\footnote{%
  Note that we do not rescale the spectral parameter $u\to\gamma u$ which is
  necessary to recover the rational Boltzmann weights describing the isotropic
  model.}
\begin{equation}
  \sinh(u)\left(  F^{2d2}_{a}  \right)^{b}_{c}
  =\left(-1\right)^{\delta_{c,-}}
   W_{\gamma=0}\bolw{a}{b}{c}{d}{u}\,
   \left(-1\right)^{\delta_{d,-}}\,.
\end{equation} 
For periodic boundary conditions the gauge factors cancel which implies that
the topological charge can be obtained from the transfer matrix in the limit
of infinite spectral parameter as
\begin{equation}
  \label{topoch0}
  \lim_{u\to\infty} \tilde\tau_{\gamma=0}(u)\,\mathrm{e}^{-Lu} = Y\,.
\end{equation}
Therefore $Y$ is part of the hierarchy of commuting integrals, i.e.\
$[\tilde{\tau}_{\gamma=0}(u),Y]=0$.  The eigenvalues of the topological charge
(which, according to (\ref{topoch0}), determine the asymptotic behaviour of
the eigenvalues of $\tilde{\tau}_{\gamma=0}$) take values from the spectrum of
the adjacency matrix (\ref{d3-adjmat}) of the anyon model
\cite{FrZu90,KlPe92}.  For the $D_3$ anyon model the eigenvalues of $Y$ are
$2$, $-1$, and $0$.  The dimension of the corresponding eigenspaces is
\begin{equation}
  \dim(\mathcal{H}_Y) = \begin{cases}
    \frac13(2^{L-1}+ (-1)^{L}) & \mathrm{for~} Y=2\\
    \frac{2}{3}(2^{L-1}+(-1)^{L}) & \mathrm{for~} Y=-1\\
    2^{L-1} &  \mathrm{for~} Y=0
  \end{cases}\,.
\end{equation}

For $\gamma\ne0$ the topological charge is not easily obtained from the
corresponding transfer matrix of the anyon model.  It is easily checked,
however, that the Boltzmann weights (\ref{bweights}) together with
$W_{\gamma=0}(u)$ satisfy the following equations
\begin{equation}
\begin{aligned}
  &\sum _{g}
  W_{\gamma=0}\bolw{f} {a}{e}{g}{u-v} 
  W_{\gamma=0}\bolw{a} {b}{g}{c}{v} W\bolw{g} {c}{e}{d}{u}\,\\
&\quad =\sum _{g^{\prime}}
  W\bolw{a} {b}{f}{g^{\prime}}{u} W_{\gamma=0}\bolw{f} {g^{\prime}}{e}{d}{v}   
  W_{\gamma=0}\bolw{b} {c}{g^{\prime}}{d}{u-v}\,.
  \end{aligned}
\end{equation}
Together with (\ref{topoch0}) this equation can be used to show that $Y$
commutes with the transfer matrix of the anyon model for arbitrary $\gamma$ as
well as the parity operator (\ref{z2par})
\begin{equation}
  [\tilde\tau(u),Y] = 0\,,\qquad[\hat{\sigma},Y]=0.
\end{equation}
Calculation of the transfer matrix eigenvalues for small $L$ we find their
large-$u$ asymptotics in the different topological sectors to be
\begin{equation}
  \label{topochg}
  \lim_{u\to\infty}\tilde\tau(u)\,\mathrm{e}^{-Lu} = \begin{cases}
    \mathrm{e}^{i\gamma(L-M)}+\mathrm{e}^{i\gamma M} 
       + O\left(\mathrm{e}^{-2u}\right)& \mathrm{for~} Y=2\\
    \omega\,\mathrm{e}^{i\gamma(L-M)}+\omega^{-1}\,\mathrm{e}^{i\gamma M} 
       + O\left(\mathrm{e}^{-2u}\right)& \mathrm{for~} Y=-1\\ 
    O\left(\mathrm{e}^{-u}\right) & \mathrm{for~} Y=0
  \end{cases}\,.
\end{equation}
Here the integer $M$ takes values such that $L-2M$ is an even (odd) multiple
of $3$ for anyon chains of even (odd) length, $\omega$ is a primitive cube
root of unity. Similarly, we observe that the parity of the transfer matrix
eigenstates is connected to the corresponding topological charge:
it is found to be $\sigma=(-1)^{L}$ in the sectors $Y=2,-1$ and
$\sigma=-(-1)^{L}$ for $Y=0$.


\section{Solution of the spectral problem}
As a consequence of the Yang-Baxter equation for the local Boltzmann weights
the $D_3$ anyon chain is integrable.  Unlike the situation for vertex models
such as the six-vertex model and the related XXZ (or $D_3$) spin chain the
spectral problem for the anyon chain can not be solved by means of the
algebraic Bethe ansatz.
Instead we will follow Refs.~\cite{FrKa14,FrKa15} and derive exact inversion
identities satisfied by the transfer matrix of inhomogeneous generalizations
of the $D_3$ anyon chain to derive Bethe equations for its spectrum.

\subsection{Inhomogeneous Chain and Inversion Identities}
Since the Boltzmann weights satisfy the Yang-Baxter equation (\ref{ybe-irf})
with the difference property the transfer matrix of a model with generic
inhomogeneities $\{y_k\}_{k=1}^L$
\begin{equation}
\label{tau_inh}
\begin{tikzpicture}

\node at (0,-1) {{$\tilde\tau(u|\{y_k\})=$}};

\draw (1,-1.5)--(5.8,-1.5);
\draw (1,-0.5)--(5.8,-0.5);

\foreach \x in  {1,2,3,4}
{
\draw (1.5*\x-0.5,-0.5)--(1.5*\x-0.5,-1.5);
\node [above] at (1.5*\x-0.5,-0.5) {$a_{\x}$};
\node [below] at (1.5*\x-0.5,-1.5) {$b_{\x}$};
}

\foreach \x in {1,2,3}
\node at (1.5*\x+0.25,-1)  {{$u-y_{\x}$}};

\draw (6.7,-0.5)--(8.5,-0.5)--(8.5,-1.5)--(6.7,-1.5);
\draw  (7.0,-1.5)--(7.0,-0.5);

\node at (7.75,-1)   {{$u-y_{L}$}};
\node [above] at (7.0,-0.5) {$a_{L}$};
\node [above] at (8.5,-0.5) {$a_{1}$};

\node [below] at (7.0,-1.5) {$b_{L}$};
\node [below] at (8.5,-1.5) {$b_{1}$};

\node at (6.25,-1.0) {$\cdots$};
\end{tikzpicture}
\end{equation}
also commutes for different spectral parameters, $\left[\tilde\tau(u|\{y_k\}),
  \tilde\tau(v|\{y_k\})\right]=0$.
Consider the following product of these transfer matrices:
\begin{equation}
\begin{tikzpicture}
\def \shift {0.5}
\def \hightu {2}

\node at (0,0)
{$\tilde\tau(u-i\gamma)\,\tilde\tau(u)=\sum\limits_{c}$}; 
\foreach \x in {1,2,4}
{   
   \draw (\shift+\x*\hightu,0)--(\shift+\x*\hightu,0.5*\hightu)--(\shift+\x*\hightu+\hightu,0.5*\hightu) -- (\shift+\x*\hightu+\hightu,-0.5*\hightu)--(\shift+\x*\hightu,-0.5*\hightu)--(\shift+\x*\hightu,0) -- (\shift+\x*\hightu+\hightu,0);  
 } 

\foreach \x in {1,2}
{
\node [above] at (\shift +\x*\hightu,0.5*\hightu) {$a_{\x}$};
\node [below] at (\shift +\x*\hightu,-0.5*\hightu) {$b_{\x}$};
\node [above right] at (\shift +\x*\hightu-0.1,-0.1) {$c_{\x}$};
\node at (\shift +\x*\hightu+0.5*\hightu , -0.25*\hightu) {{$u-y_{\x}-i\gamma$}};
\node at (\shift +\x*\hightu+0.5*\hightu,0.25*\hightu) {{$u-y_{\x}$}};
} 

\node at (\shift +4*\hightu+0.5*\hightu , -0.25*\hightu) {{$u-y_{L}-i\gamma$}};
\node at (\shift +4*\hightu+0.5*\hightu,0.25*\hightu) {{$u-y_{L}$}};

\node [above] at (\shift +3*\hightu,0.5*\hightu) {$a_{3}$};
\node [below] at (\shift +3*\hightu,-0.5*\hightu) {$b_{3}$};
\node [above right] at (\shift +3*\hightu-0.1,-0.1) {$c_{3}$};

\node [above] at (\shift +4*\hightu,0.5*\hightu) {$a_{L}$};
\node [below] at (\shift +4*\hightu,-0.5*\hightu) {$b_{L}$};
\node [above right] at (\shift +4*\hightu-0.1,-0.1) {$c_{L}$};

\node [above] at (\shift +5*\hightu,0.5*\hightu) {$a_{1}$};
\node [below] at (\shift +5*\hightu,-0.5*\hightu) {$b_{1}$};
\node [above right] at (\shift +5*\hightu-0.1,-0.1) {$c_{1}$};

\draw  (\shift+3*\hightu,0)--(\shift+3*\hightu+0.3,0);
\draw  (\shift+4*\hightu,0)--(\shift+4*\hightu-0.3,0);
\draw [dashed] (\shift+3*\hightu+0.3,0)--(\shift+4*\hightu-0.3,0);

\draw  (\shift+3*\hightu,0.5*\hightu)--(\shift+3*\hightu+0.3,0.5*\hightu);
\draw  (\shift+4*\hightu,0.5*\hightu)--(\shift+4*\hightu-0.3,0.5*\hightu);
\draw [dashed] (\shift+3*\hightu+0.3,0.5*\hightu)--(\shift+4*\hightu-0.3,0.5*\hightu);

\draw  (\shift+3*\hightu,-0.5*\hightu)--(\shift+3*\hightu+0.3,-0.5*\hightu);
\draw  (\shift+4*\hightu,-0.5*\hightu)--(\shift+4*\hightu-0.3,-0.5*\hightu);
\draw [dashed] (\shift+3*\hightu+0.3,-0.5*\hightu)--(\shift+4*\hightu-0.3,-0.5*\hightu);
\end{tikzpicture}
\end{equation}
where the summation is over the possible anyon labels $\{c_j\}_{j=1}^L$ on the
inner nodes.  
Choosing the spectral parameter $u$ to take the value $y_{k}$ from the set of
inhomogeneities we can use the initial condition (\ref{id:initval}) of the
Boltzmann weights together with crossing and unitarity (\ref{id:crossing}),
(\ref{id:unitarity}) to obtain
\begin{equation}
\begin{aligned}
&
\begin{tikzpicture}

\def \shiftq {-0.3}
\def\shift {\shiftq+1}								\def \shiftb {\shiftq+3*\hightu+2.5}
\def \shiftc {-1.7}
\def \hightu {2} 

\node at (0,-0.1) {$\sum\limits_{c_k,c_{k+1}}$};
\draw
(\shiftq+\hightu,0)-- ++(0,0.5*\hightu) -- ++(\hightu,0)
  -- ++(0,-\hightu) -- ++(-\hightu,0) -- ++(0,0.5*\hightu)
  -- ++(\hightu,0); 
\draw
(\shiftq+2*\hightu,0)-- ++(0,0.5*\hightu) -- ++(\hightu,0)
  -- ++(0,-\hightu) -- ++(-\hightu,0) -- ++(0,0.5*\hightu)
  -- ++(\hightu,0); 

\draw [dashed]  (\shiftq+\hightu,0)-- +(-0.5,0);
\draw [dashed]  (\shiftq+\hightu,0.5*\hightu) -- +(-0.5,0); %
\draw [dashed]  (\shiftq+\hightu,-0.5*\hightu) -- +(-0.5,0); %
\draw [dashed]  (\shiftq+3*\hightu,0)-- +(0.5,0);
\draw [dashed]  (\shiftq+3*\hightu,0.5*\hightu) -- +(0.5,0); %
\draw [dashed]  (\shiftq+3*\hightu,-0.5*\hightu) -- +(0.5,0); %

\node[above] at (\shiftq+\hightu, 0.5*\hightu) {$a_{k}$};
\node[above] at (\shiftq+2*\hightu,0.5*\hightu) {$a_{k+1}$};
\node[above] at (\shiftq+3*\hightu,0.5*\hightu) {$a_{k+2}$};

\node[below] at (\shiftq+\hightu, -0.5*\hightu) {$b_{k}$};
\node[below] at (\shiftq+2*\hightu,-0.5*\hightu) {$b_{k+1}$};
\node[below] at (\shiftq+3*\hightu,-0.5*\hightu) {$b_{k+2}$};

\node [above left] at (\shiftq+\hightu+0.1,-0.1) {$c_{k}$};
\node [above right] at (\shiftq+2*\hightu-0.1,-0.1) {$c_{k+1}$};
\node [above right] at (\shiftq+3*\hightu-0.1,-0.1) {$c_{k+2}$};

\node at (\shiftq +\hightu+0.5*\hightu , 0.3*\hightu) {$0$};
\node at (\shiftq+\hightu+0.5*\hightu , -0.3*\hightu) {$-i\gamma$};
\node at (\shiftq +\hightu+1.5*\hightu , 0.3*\hightu) {$\Delta_{k,k+1}$};
\node at (\shiftq+\hightu+1.5*\hightu , -0.3*\hightu) {$\Delta_{k,k+1}-i\gamma$};

\end{tikzpicture}
\\
& \qquad
\begin{tikzpicture}

\def \shiftq {-1.0}
\def\shift {\shiftq+1}
\def \shiftb {\shiftq+3*\hightu+2.5}
\def \shiftc {-1.7}
\def \hightu {2} 

\node at (\shiftq-0.5,-0.1)
  {$=$};
\draw (\shiftq+0.4*\hightu,0.5*\hightu) -- ++(0,-\hightu); 
\draw [dashed]  (\shiftq+0.4*\hightu,0)-- +(-0.5,0);
\draw [dashed]  (\shiftq+0.4*\hightu,0.5*\hightu) -- +(-0.5,0); %
\draw [dashed]  (\shiftq+0.4*\hightu,-0.5*\hightu) -- +(-0.5,0); %

\node[above] at (\shiftq+0.4*\hightu, 0.5*\hightu) {$a_{k}$};
\node[below] at (\shiftq+0.4*\hightu, -0.5*\hightu) {$b_{k}$};
\node [above left] at (\shiftq+0.4*\hightu+0.1,-0.1) {$a_{k+1}$};
\node [right] at (\shiftq+0.6*\hightu,-0.1)
  {$\mathcal{G}_{[b_ka_{k+1}]}^{-1}\,\delta_{b_{k+1}}^{\bar{a}_{k+1}}\, 
    \sum\limits_{c_{k+1}} \mathcal{G}_{[a_{k+1}c_{k+1}]}$};

\draw (\shiftq+3.25*\hightu,0)-- ++(0,0.5*\hightu) -- ++(\hightu,0)
  -- ++(0,-\hightu) -- ++(-\hightu,0) -- ++(0,0.5*\hightu)
  -- ++(\hightu,0); 
\draw [dashed]  (\shiftq+4.25*\hightu,0)-- +(0.5,0);
\draw [dashed]  (\shiftq+4.25*\hightu,0.5*\hightu) -- +(0.5,0); 
\draw [dashed]  (\shiftq+4.25*\hightu,-0.5*\hightu) -- +(0.5,0); 

\node[above] at (\shiftq+3.25*\hightu, 0.5*\hightu) {$a_{k+1}$};
\node[below] at (\shiftq+3.25*\hightu, -0.5*\hightu) {$\bar{a}_{k+1}$};
\node [above right] at (\shiftq+3.25*\hightu-0.1,-0.1) {$c_{k+1}$};
\node at (\shiftq +3.75*\hightu, 0.3*\hightu) {$\Delta_{k,k+1}$};
\node at (\shiftq +3.75*\hightu, -0.3*\hightu) {$\Delta_{k,k+1}-i\gamma$};
\node[above] at (\shiftq+4.25*\hightu,0.5*\hightu) {$a_{k+2}$};
\node[below] at (\shiftq+4.25*\hightu,-0.5*\hightu) {$b_{k+2}$};
\node [above right] at (\shiftq+4.25*\hightu-0.1,-0.1) {$c_{k+2}$};

\node[right] at (\shiftq+4.75*\hightu,-0.1) {$\times(\sinh(i\gamma))^2$};
\end{tikzpicture}
\\
& \qquad
\begin{tikzpicture}

\def \shiftq {-1.0}
\def\shift {\shiftq+1}
\def \shiftb {\shiftq+3*\hightu+2.5}
\def \shiftc {-1.7}
\def \hightu {2} 

\node at (\shiftq-0.5,-0.1)
  {$=$};
\draw (\shiftq+0.4*\hightu,0.5*\hightu) -- ++(0,-\hightu); 
\draw [dashed]  (\shiftq+0.4*\hightu,0)-- +(-0.5,0);
\draw [dashed]  (\shiftq+0.4*\hightu,0.5*\hightu) -- +(-0.5,0); %
\draw [dashed]  (\shiftq+0.4*\hightu,-0.5*\hightu) -- +(-0.5,0); %

\node[above] at (\shiftq+0.4*\hightu, 0.5*\hightu) {$a_{k}$};
\node[below] at (\shiftq+0.4*\hightu, -0.5*\hightu) {$b_{k}$};
\node [above left] at (\shiftq+0.4*\hightu+0.1,-0.1) {$a_{k+1}$};
\node [right] at (\shiftq+0.6*\hightu,-0.1)
  {$\mathcal{G}_{[b_ka_{k+1}]}^{-1}\,\delta_{b_{k+1}}^{\bar{a}_{k+1}}\,
    \delta_{b_{k+2}}^{\bar{a}_{k+2}}\,
    \mathcal{G}_{[a_{k+2}c_{k+2}]}$};

\draw (\shiftq+3.25*\hightu,0.5*\hightu) -- ++(0,-\hightu); 
\draw [dashed]  (\shiftq+3.25*\hightu,0)-- +(0.5,0);
\draw [dashed]  (\shiftq+3.25*\hightu,0.5*\hightu) -- +(0.5,0); %
\draw [dashed]  (\shiftq+3.25*\hightu,-0.5*\hightu) -- +(0.5,0); %

\node[above] at (\shiftq+3.25*\hightu,0.5*\hightu) {$a_{k+2}$};
\node[below] at (\shiftq+3.25*\hightu,-0.5*\hightu) {$\bar{a}_{k+2}$};
\node [above right] at (\shiftq+3.25*\hightu-0.1,-0.1) {$c_{k+2}$};

\node [right] at (\shiftq+4*\hightu-0.1,-0.1) {$\times$};
\end{tikzpicture}
\\
&\qquad\qquad\qquad \times(\sinh(i\gamma))^2\,
  \rho(\Delta_{k,k+1})\rho(-\Delta_{k,k+1})
\end{aligned}
\end{equation}
with $\Delta_{k\ell}=y_k-y_\ell$.  Iterating this operation we obtain the
following set of discrete inversion identities satisfied by the transfer
matrix
\begin{equation}
  \label{invid}
  \tilde\tau(y_{k}-i\gamma)\,\tilde\tau(y_{k}) = \hat{\sigma}\,
  \prod\limits_{\ell=1}^{L}\rho(y_k-y_\ell)\rho(y_\ell-y_k)\,,
  \qquad k=1,\ldots,L\,.
\end{equation}
Periodic boundary conditions imply
\begin{equation}
  \prod_{\ell=1}^L \tilde\tau(y_k) = \prod_{k,\ell=1}^L \rho(y_k-y_\ell)\,
  \mathbf{1}\,,
  \qquad
  \prod_{\ell=1}^L \tilde\tau(y_k-i\gamma) = \hat\sigma^L\,\prod_{k,\ell=1}^L
  \rho(y_k-y_\ell)\,.
\end{equation}
Therefore only $L-1$ of the identities (\ref{invid}) are independent.

As a consequence of the commutativity of the transfer matrices and the parity
operator $\hat{\sigma}$, the eigenvalues of $\tilde\tau(u|\{y_k\})$ satisfy
similar identities, i.e.\
\begin{equation}
  \label{invid-ev}
  \Lambda(y_{k}-i\gamma)\,\Lambda(y_{k}) =
  \pm(-1)^L \prod\limits_{\ell=1}^{L}\rho(y_k-y_\ell)\rho(y_\ell-y_k)\,,
  \qquad k=1,\ldots,L-1\,,
\end{equation}
where the sign $+$ ($-$) has to be chosen for the topological sectors $Y=2,-1$
($Y=0$).  Eqs.~(\ref{invid}) or (\ref{invid-ev}) together with information on
the analytical properties of the transfer matrix allow to compute its
eigenvalues: By construction they are Fourier polynomials in $\mathrm{e}^{2u}$
of degree $\le L/2$.  The coefficients of the leading terms $\mathrm{e}^{\pm
  Lu}$ are determined by the topological charge according to
(\ref{topochg}).  The remaining $L-1$ coefficients can be computed by solving
the quadratic equations (\ref{invid-ev}). 

\subsection{$TQ$-equation for the $D_3$ anyon chain}
For an efficient calculation of transfer matrix eigenvalues for large systems
the inversion identities are not suitable.  Note, however, that for a generic
choice of inhomogeneities, i.e.\ $y_k-y_\ell\ne 0,\pm i\gamma$ for $k\ne\ell$
they are formally equivalent to a $TQ$-type equation
\begin{equation}
  \label{tq-eq}
  \Lambda(u) q(u) = \Delta_+(u) q(u-i\gamma) + \Delta_-(u) q(u+i\gamma)
\end{equation}
restricted to the discrete set of points $u\in\{y_k,y_k-i\gamma\}$ provided
that the functions $\Delta_\pm$ factorize the right hand side of
(\ref{invid-ev}) as
\begin{equation}
  \Delta_+(y_k)\Delta_-(y_k-i\gamma) = 
    \pm(-1)^L \prod\limits_{\ell=1}^{L}\rho(y_k-y_\ell)\rho(y_\ell-y_k)\,.
\end{equation}
and satisfy  $\Delta_+(y_k-i\gamma) = \Delta_-(y_k) = 0$.  In the present case
these conditions are met by the choice
\begin{equation}
  \label{tq-fac}
  \Delta_+(u) = \omega_+ \prod_{\ell=1}^L \sinh(u-y_\ell+i\gamma)\,,
  \qquad
  \Delta_-(u) = \omega_- \prod_{\ell=1}^L \sinh(u-y_\ell)\,,
\end{equation}
provided that $\omega_+\omega_-=1$ ($-1)$ in the topological sectors $Y=2,-1$
($Y=0$). 

$TQ$-equations such as (\ref{tq-eq}) which are valid for arbitrary
$u\in\mathbb{C}$ are obtained in the Bethe ansatz formulations of the spectral
problem of integrable systems such as the six-vertex model \cite{Baxter:book}.
Provided that they allow for a sufficiently simple (e.g.\ polynomial) ansatz
for the functions $q(u)$ they can be solved using the Bethe equations for the
finitely many zeroes of these functions.

\subsubsection{Sectors $Y=2,-1$}
We take the Fourier polynomial 
\begin{equation}
  \label{qfunc2-1}
  q(u) = \prod_{k=1}^M\sinh\left(u-u_k+\frac{i}{2}\gamma\right)\,,
\end{equation}
parametrized by $M$ complex numbers $u_k$ as ansatz for the $q$-functions in
these sectors.  
The large $u$ asymptotics of the corresponding transfer matrix eigenvalues
$\Lambda(u)$ as obtained from the $TQ$-equation (\ref{tq-eq}) together with
the factorization (\ref{tq-fac}) of $\Lambda(y_k-i\gamma)\Lambda(y_k)$,
\begin{equation}
  \label{asy2-1}
  \lim_{u\to\infty} \Lambda(u) \mathrm{e}^{-Lu+\sum_{\ell=1}^L y_\ell}
  = \omega_+\mathrm{e}^{i\gamma(L-M)} + \omega_-\mathrm{e}^{i\gamma M}\,,
\end{equation}
agrees with (\ref{topochg}) provided that $L-2M$ is an even (odd) multiple of
$3$ for even (odd) $L$ and $\omega_+=1/\omega_-$ is $1$ in the topological
sector $Y=2$, and $\omega$, a primitive cube root of unity, for $Y=-1$.
Analyticity of
\begin{equation}
  \label{tmev2-1}
  \Lambda^{(Y)}(u)= \Delta_+^{(Y)}(u) \prod_{k=1}^M
  \frac{\sinh\left(u-u_k-\frac{i}{2}\gamma\right)}{
        \sinh\left(u-u_k+\frac{i}{2}\gamma\right)}
  +\Delta_-^{(Y)}(u) \prod_{k=1}^M
  \frac{\sinh\left(u-u_k+\frac{3i}{2}\gamma\right)}{
        \sinh\left(u-u_k+\frac{i}{2}\gamma\right)}
\end{equation}
implies that the parameters $u_k$ have to satisfy the Bethe equations
\begin{equation}
  \label{bae2-1}
  \frac{\omega_+}{\omega_-}\,
  \prod_{\ell=1}^L\frac{\sinh(u_k-y_\ell+\frac{i}{2}\gamma)}{
                      \sinh(u_k-y_\ell-\frac{i}{2}\gamma)} = -
  \prod_{j=1}^M\frac{\sinh(u_k-u_j+i\gamma)}{\sinh(u_k-u_j-i\gamma)}\,,
  \quad k=1,\ldots,M\,.
\end{equation}
These equations coincide with those for the spin-1/2 XXZ (or $D_3$) spin chain
for periodic (in the sector $Y=2$) or twisted boundary conditions with
(diagonal) twist $\omega_+/\omega_-=\omega^2$ for (for $Y=-1$).  For the spin
chain with these boundary conditions the number of Bethe roots $m$ is related
to the conserved $U(1)$ charge, i.e.\ the magnetization $S^z=L/2-M$.  A
particularly simple eigenstate of the XXZ model is the completely polarized
state corresponding to $M=0$.  In the anyon chain such a state can be realized
for chains of length $L$ being multiples of $3$.  The corresponding
eigenstates are in the subspace spanned by the fusion path states
$\ket{a_122a_4\ldots22a_{L-2}22}$ and its translations
\cite{GATH13}.\footnote{%
  Note that in the critical disordered regime ($\gamma\in\mathbb{R}$)
  considered here these states are never ground states of the anyon chain.}
The $a_i$ represent the state $\ket{a_i}=(\ket{+}-\ket{-})/\sqrt{2}$ at
position $i$.  The translationally invariant zero momentum superposition
belongs to the sector with topological charge $Y=2$, the linear combinations
with momentum $\pm 2\pi/3$ appear in the sector $Y=-1$.

Similarly, the fusion path states contributing to transfer matrix eigenstates
with $M>0$ (appearing for chain of length $L$ with $L-2M$ being multiples of
$3$) can be built by concatenation of $L-2M$ segments $\ket{a22}$ and, in
addition, $M$ two-site segments
\begin{equation}
  \label{any-magnon}
  \ket{b2} = \frac1{\sqrt{2}}\left(\ket{+2}+\ket{-2}\right)\,
  \quad\mathrm{or}\quad\ket{22}\,.
\end{equation}

\subsubsection{Sector $Y=0$}
In the topological sector with $Y=0$ the ansatz (\ref{qfunc2-1}) for the
$q$-functions does not reproduce the asymptotic behaviour (\ref{topochg}) of
the transfer matrix eigenvalues.  Instead we choose
\begin{equation}
  \label{qfunc-0}
  q(u) = \prod_{k=1}^{\widetilde{M}}
  \sinh\frac12\left(u-u_k+\frac{i}{2}\gamma\right)\,.  
\end{equation}
Again, the large-$u$ behaviour as obtained from the $TQ$-equation
(\ref{tq-eq}) 
\begin{equation}
  \label{asy-0}
  \lim_{u\to\infty} \Lambda(u)\mathrm{e}^{-Lu+\sum_\ell y_\ell} = 
  \omega_+ \mathrm{e}^{i\gamma\left(L-\widetilde{M}/2\right)} + 
  \omega_- \mathrm{e}^{i\gamma\widetilde{M}/2}
\end{equation}
has to match (\ref{topochg}).  This implies $\omega_\pm=\pm1$ and
$\widetilde{M}=L$.  As before, requiring analyticity of the transfer matrix
eigenvalues
\begin{equation}
  \label{tmev-0}
  \begin{aligned}
    \Lambda^{(Y=0)}(u) =& \prod_{\ell=1}^L \sinh(u -y_\ell +  i\gamma)\,
    \prod_{k=1}^{L }\frac{  \sinh\frac{1}{2}\left(u-u_{k}-\frac{i
        }{2} \gamma  \right)  }{
      \sinh\frac{1}{2}\left(u-u_{k}+\frac{i }{2} \gamma  \right)   }\\
    & \qquad-
    \prod_{\ell=1}^L \sinh(u-y_\ell)^{L}\, \prod_{i=1}^{L}  \frac{  \sinh
      \frac{1}{2}\left(u-u_{k}+\frac{3i}{2}  \gamma  \right)  }{   \sinh
      \frac{1}{2}\left(u-u_{k}+\frac{i }{2}\gamma   \right)   } \,,
  \end{aligned}
\end{equation}
leads to the Bethe equations determining the parameters $u_k$ in the
$q$-function (\ref{qfunc-0}) for the topological sector $Y=0$:
\begin{equation}
  \label{bae-0}
  \prod_{\ell=1}^L\frac{\sinh(u_k-y_\ell+\frac{i}{2}\gamma)}{
                      \sinh(u_k-y_\ell-\frac{i}{2}\gamma)} = 
  \prod_{j=1}^L\frac{\sinh\frac12\left(u_k-u_j+i\gamma\right)}{
    \sinh\frac12\left(u_k-u_j-i\gamma\right)}\,,
  \quad k=1,\ldots,L\,.
\end{equation}
The eigenvalues (\ref{tmev-0}) coincide with those of the XXZ spin-$1/2$ chain
with antiperiodic boundary conditions \cite{BBOY95,NiWF09},
\begin{equation}
  \label{xxztwist}
  \sigma^\alpha_{L+1} =K^{-1}\sigma^\alpha_1K\,, 
  \qquad K=\left(\begin{array}{cc}
      0&1\\1&0\end{array}\right)\,.
\end{equation}

\section{Ground state and low lying excitations}
To find the spectrum of the anyon chain the Bethe equations (\ref{bae2-1}),
(\ref{bae-0}) have to be solved for the different topological sectors.  From
Eqs.~(\ref{tmev2-1}) and (\ref{tmev-0}) the eigenvalues of the Hamiltonian
(\ref{irf:hamil}) corresponding to a set $\{u_k\}$ of Bethe roots are
\begin{equation}
\begin{aligned}
  &Y=2,-1:
  &&E(\{u_k\}) = L\cos\gamma
     - 2\sum_{k=1}^M \frac{\sin^2\gamma}{\cosh(2u_k)-\cos\gamma}\,,\\
  &Y=0:
  &&E(\{u_k\}) = L\cos\gamma -  \sum_{k=1}^L 
        \frac{\sin\gamma\,\sin\frac{\gamma}{2}}{\cosh u_k -
          \cos\frac{\gamma}{2}}\,. 
\end{aligned}
\end{equation}

\subsection{Thermodynamic Limit}
The thermodynamic properties of the system do not depend on the boundary
conditions.  Since we have identified the spectrum of the $D_3$ anyon chain
with that of the XXZ Heisenberg spin chain with boundary conditions depending
on the topological sector the structure of the phase diagram as well as
certain bulk properties can be infered from the corresponding properties of
the XXZ chain.  For $J^z/|J|<-1$ ($\ge+1$) the latter is a gapped ferromagnet
(antferromagnet).
The simple fusion path states in the $M=0$ sectors for topological charge
$Y=2,-1$ discussed above form the ground states of the anyon model in the
former interval and lead to the $\mathbb{Z}_3$-sublattice structure in this
phase observed in Ref.~\cite{GATH13}.  Similarly, the ground state for
$J^z/|J|\ge1$ is in the $M=L/2$ sector implying a $\mathbb{Z}_2$-sublattice
structure as in the antiferromagnetic XXZ chain.

For $-1<J^z/|J|\le1$ (or $0\le\gamma<\pi$ in the present notation) the
XXZ model is in its critical regime and known to be conformally
invariant.  This observation confirms the conclusion on the location
of the phase boundaries reached from the numerical analysis of the
anyon chain in Ref.~\cite{GATH13}.
The ground state of the spin chain in the critical phase is in the zero
magnetization sector $M=L/2$.  The energy density has been obtained within the
root density formalism applied to the Bethe equations of the XXZ spin-1/2
chain (with $J=+1$) yielding \cite{YaYa66b}
\begin{equation}
  \epsilon_\infty = \lim_{L\to\infty} \frac{E_0(L)}{L}
  = \cos\gamma -\frac{\sin\gamma}{\gamma} \int_{-\infty}^\infty \mathrm{d}u\,
         \frac{\sin\gamma}{\cosh\left({\pi}u/{\gamma}\right)
                           \left(\cosh(2u)-\cos\gamma\right)}\,.
\end{equation}
The elementary excitations have a linear dispersion with Fermi velocity
\begin{equation}
  v_F = \pi\frac{\sin\gamma}{\gamma}\,.
\end{equation}
In the spectrum of the $D_3$ anyon chain with $-1<J^z\le1$ (and $J=+1$) these
massless excitations are realized near momenta $k\,\pi/2$, $k=0,1,2,3$.  This
is the $\mathbb{Z}_4$ critical region identified in Ref.~\cite{GATH13}.
Similar to the spin chain the spectrum of the anyon Hamiltonian is invariant
under the change $J\to-J$ for lattice sizes $L$ being multiples of four, i.e.\
\begin{equation}
  \mathrm{spec}\left(\tilde{H}(J,J^z)\right) = 
  \mathrm{spec}\left(\tilde{H}(-J,J^z)\right)\,.
\end{equation}
The corresponding unitary transformation relating the two Hamiltonians changes
the momentum such that gapless excitations for $J=-1$ exist at $0$, $\pi$
only, characteristic for the $\mathbb{Z}_2$ critical region of the $D_3$ anyon
chain.

\subsection{Finite size spectrum and conformal field theory of the Heisenberg
  spin chain}
The low energy excitations of the periodic spin chain (\ref{ham-xxz}) for
$-1<J^z/|J|\le 1$ are described by a conformal field theory with central
charge $c=1$ \cite{AvDo86,Hame86}.  The operator content of this CFT can be
identified from the finite size scaling of the energies of the ground state
and excitations through \cite{BlCN86,Affl86,Card86a}
\begin{equation}
  \label{cft_fs}
  E_{h\bar{h}}(L) - L\epsilon_\infty = -\frac{\pi v_F}{6L}\,c
    + \frac{2\pi v_F}{L}\left(h+\bar{h}
      + R_{h\bar{h}}(L)
    \right)\,.
\end{equation}
Here $(h,\bar{h})$ are the conformal weights of primary operators of the CFT,
$R_{h\bar{h}}(L)$ subsumes the subleading corrections to scaling of the
corresponding energy level from which the irrelevant perturbations of the
conformal fixed point Hamiltonian present in the lattice model can be
identified.

For periodic boundary conditions the low energy spectrum of the XXZ chain is
that of a Gaussian model, i.e.\ a free boson with compactification radius
$r=r_G(\gamma)=\sqrt{(\pi-\gamma)/(2\pi)} \in(0,1/\sqrt{2}]$ for exchange
constants $(J,J^z)=(\pm1,\cos\gamma)$ in (\ref{ham-xxz}) \cite{AlBB88}.  The
scaling dimensions of the primary operators, highest weights of two commuting
$U(1)$ Kac-Moody algebras, are
\begin{equation}
  \label{xxzper}
  X_{mn}(r) = h_{mn}(r)+\bar{h}_{mn}(r) = {r^2 m^2} +
  \frac{n^2}{4r^2}\,.
\end{equation}
The quantum number $m$ is the $U(1)$ charge $\frac12 \sum\sigma_j^z$ in the
corresponding state, the vorticity $n$ is related to its momentum.
For twisted boundary conditions with a diagonal unimodular twist matrix in
(\ref{xxztwist}) parametrized by an angle $\phi$ the scaling dimensions
(\ref{xxzper}) are modified to $X_{m,n+\phi/\pi}$.
In addition there exists a marginal operator in the sector $m=0$ with
anomalous dimension $X=2$ for all $\gamma$.  Its presence results in the
continuous line $0\le\gamma<\pi$ of critical points.  The partition function
of the theory satisfies the duality condition $Z(r)=Z(1/2r)$ as a consequence
of the invariance of the spectrum (\ref{xxzper}) under the simultaneous
interchanges $m\leftrightarrow n$ and $r\leftrightarrow 1/2r$.

In the case of the anti-diagonal twisted boundary conditions (\ref{xxztwist})
the $O(2)$ bulk symmetry of the XXZ spin chain is broken up to a
$\mathbb{Z}_{2}\otimes\mathbb{Z}_{2} $ with the factors being generated by
rotation around the $z$-axis by $\pi$ and a global spin flip,
respectively. The low-energy spectrum is that of a $U(1)$-twisted Kac–Moody
algebra without conserved charge. In this case the conformal weights are
\cite{ABGR88}
\begin{equation}
  \label{xxzaper}
  (h,\bar{h})_{k_1k_2} = \left(\frac{(4k_1+1)^2}{16},
    \frac{(4k_2+1)^2}{16} \right)\,,\qquad k_1,k_2\in \mathbb{Z}\,
\end{equation}
independent of the anisotropy.

\subsection{Conformal field theory of the $D_3$ anyon chain}
%
Based on our identification of the spectrum of the $D_3$ anyon chain with
subsets of that of the XXZ spin chain with boundary conditions depending on
the topological sector we can now deduce the operator content of the low
energy field theory for the anyon model:
the finite size spectrum of low lying eigenvalues of the Hamiltonian
(\ref{irf:hamil}) in the sectors with topological charge $Y=2$ correspond to
primary operators with dimensions (\ref{xxzper}).  The selection rules on the
$U(1)$ charge $M=L/2-m$ and the twist angles $\pm 2\pi/3$ in the sector
$Y=-1$, imply that levels with
\begin{equation}
  \label{cft:select}
  2m\in3\mathbb{Z}\,\quad\mathrm{and}\quad 3n\in\mathbb{Z}\,
\end{equation}
($2m=L\mod 2$) appear in the spectrum of the anyon chain.  This part of the
spectrum is that of a Gaussian model with compactification radius
\begin{equation}
  \label{radius_any}
  \tilde{r}(\gamma) = 3\,r_G(\gamma) = 3\sqrt{\frac{\pi-\gamma}{2\pi}}\,.
\end{equation}
In addition, from the low-lying levels in the sector $Y=0$ we know that the
CFT contains operators with scaling dimension $1/8$ and $9/8$.  
Together these spectral data lead to the identification of the
underlying CFT as the $\mathbb{Z}_2$ orbifolds of a boson compactified
on a circle of radius $\tilde{r}(\gamma)$.  These CFTs are constructed
by projection onto states which are invariant under a symmetry of the
model \cite{Ginsparg88,DVVV89}: for example, the theory with
$\tilde{r}(\gamma=0)=3/\sqrt{2}$ is derived from the $SU(2)_1$
Wess-Zumino-Novikov-Witten model -- the effective field theory
describing the low energy excitations of the isotropic ($\gamma=0$)
Heisenberg chain -- by 'modding out' the discrete subgroup $D_3$ of
$SO(3)\subset SU(2)$.  The primary fields in the $\mathbb{Z}_2$
orbifold CFT are
\begin{itemize}
\item the identity corresponding to the ground state, with conformal weights
  $h_0=\bar{h_0}=0$,
\item the $\mathbb{Z}_2$ twist fields $\sigma_{i}$ and $\tau_{i}$, $i=1,2$,
  with scaling dimensions $h_\sigma+\bar{h}_\sigma=1/8$ and
  $h_\tau+\bar{h}_\tau=9/8$,
\item a field $\Theta$ with scaling dimension $2$,
\end{itemize}
and, in addition, the primaries $\tilde{\phi}_{mn}$ with scaling dimensions
$X_{mn}(\tilde{r})$ given by (\ref{xxzper}).

For rational values of $\tilde{r}^2$ these CFTs have extended symmetries
allowing for a description in terms of a \emph{finite} number of primary
fields (admissible highest weight representations with respect to the extended
symmetry algebra).
For the $\mathbb{Z}_2$ orbifold theories with $c=1$ and compactification radii
$\tilde{r}(\gamma)=\sqrt{p/2}$ (or $1/\sqrt{2p}$ by duality) with $p$ integer
these are the identity, the twist fields, the marginal field $\Theta$ and, in
addition, two degenerate fields $\Phi^{1,2}$ with conformal weight
$h_\Phi=p/4$, and $p-1$ fields $\phi_\lambda$ with conformal weight
$h_\lambda=\lambda^2/(4p)$ for $\lambda=1,2,\ldots,p-1$.

\begin{figure}[t]
\includegraphics[width=0.95\textwidth]{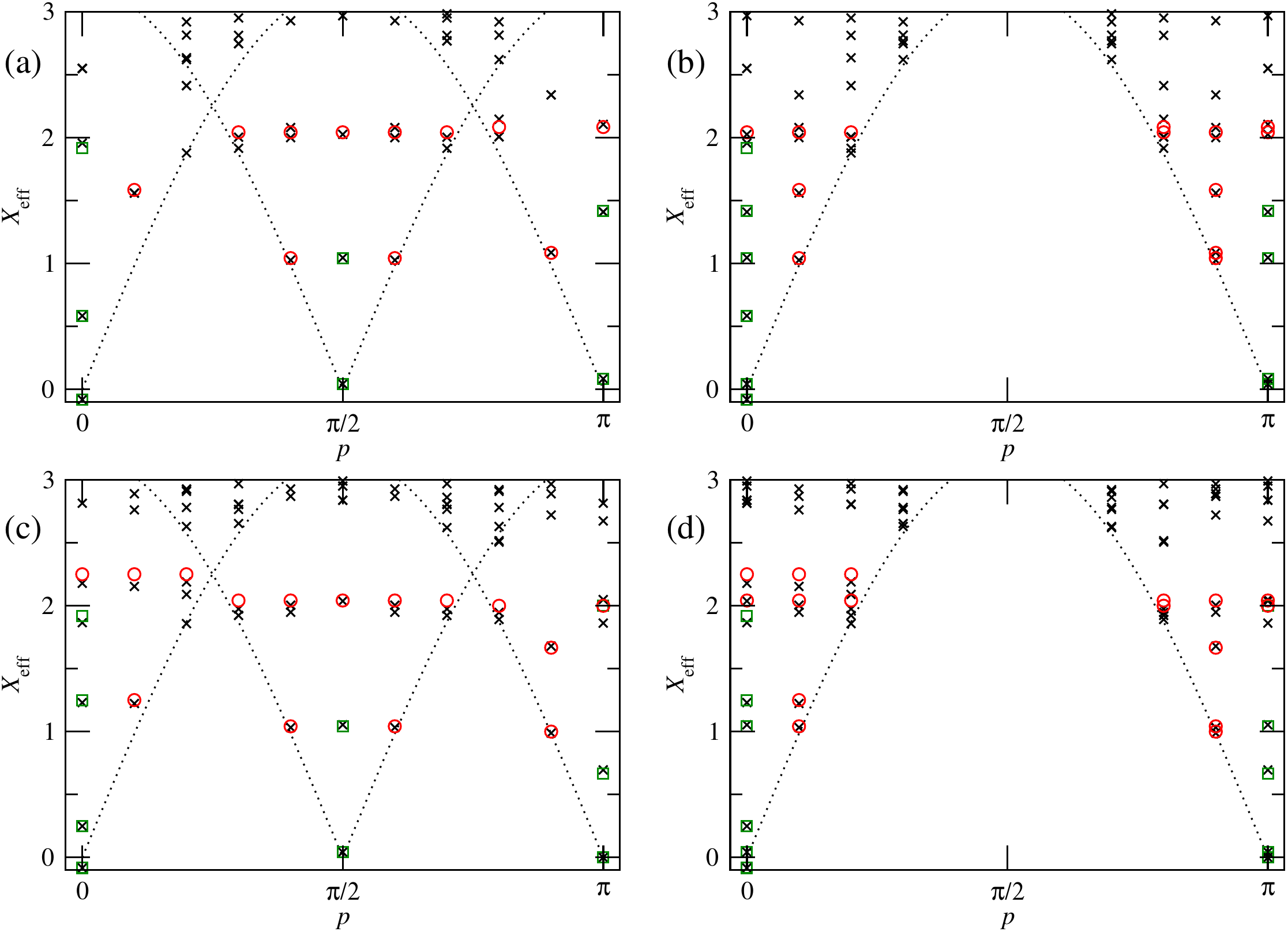}
\caption{Effective scaling dimensions
  $X_\mathrm{eff}=L\left(E-L\epsilon_\infty\right)/(2\pi v_F)$ of the $D_3$
  anyon chain vs.\ momentum of the corresponding excitations at some of the
  points (\ref{cft-orb}) in the critical region.  
  Spectra are shown for coupling constant $\gamma=2\pi/3$ in the anyon chain
  corresponding to the $\mathbb{Z}_2$ orbifold CFT with $p=3$ ($\mathbb{Z}_4$
  parafermions) in the $\mathbb{Z}_4$ critical region of the anyon chain with
  $J=+1$ (a) and the $\mathbb{Z}_2$ critical region, $J=-1$ (b). 
  Bottom panels (c) and (d): similar for $\gamma=\pi/3$ corresponding to the
  $\mathbb{Z}_2$ orbifold CFT with $p=6$ (the superconformal minimal model
  with $c=1$). 
  Symbols $\times$ are the finite size data for $L=20$ sites, green squares
  are the CFT predictions for the scaling dimensions of primary 
  fields, red circles represent some of the descendents.  Dotted lines
  indicate the properly re-scaled quasi-particle dispersion as obtained from
  the exact solution of the model, i.e.\ the lower bounds of the continua of
  collective excitations in the anyon chain. 
}
\label{fig:cftp}
\end{figure}

From (\ref{radius_any}) these rational CFTs with $p=1,\ldots,9$ are realized
for coupling constants
\begin{equation}
  \label{cft-orb}
  \gamma(p) = \pi\left(1-\frac{p}{9}\right)\,,\quad p=1,\ldots,9\,.
\end{equation}
Apart from the $SU(2)/D_3$ model for $p=9$ mentioned above, the $D_3$ anyon
model provides lattice realizations of the Kosterlitz-Thouless theory ($p=1$),
two uncoupled copies of the critical Ising model ($p=2$), $Z_4$ parafermions
($p=3$), the four-state Potts model ($p=4$) and the superconformal minimal
model ($p=6$), all with central charge $c=1$ \cite{Ginsparg88}.
The finite size spectra of the anyon chain for the $\mathbb{Z}_4$ parafermion
CFT, $p=3$ or $\gamma=2\pi/3$, and the superconformal minimal model, $p=6$ or
$\gamma=\pi/3$, are shown in Figure~\ref{fig:cftp}, both for the
$\mathbb{Z}_4$ ($J=+1$) and the $\mathbb{Z}_2$ critical case ($J=-1$).

Estimates for the location of the points $p=1,\ldots,8$ in the critical region
have been given in Ref.~\cite{GATH13} based on the numerical analysis of the
finite size spectrum of the $su(2)_4$ spin-$1$ anyon chain.  In their
discussion of the phase diagram the authors of that paper have also linked the
transition from the critical regime into the gapped phase with $\mathbb{Z}_2$
sublattice symmetry with the appearance of a marginal operator with zero
momentum in the topological sector $Y=2$ of the ground state for $p=9$: in
fact, it is well known from the analysis of the subleading corrections to
scaling $R_{h\bar{h}}(L)$ in (\ref{cft_fs}) for the XXZ spin chain that its
continuum limit differs from the conformally invariant fixed point Hamiltonian
by terms involving an operator from the conformal block of the identity with
scaling dimension $X=4$ and the primary operator with dimension
$X=2\pi/(\pi-\gamma)$ \cite{AlBB88,NiWF09}.  In the $\mathbb{Z}_2$ orbifold
CFT with compactification radius $\tilde{r}$ for the anyon chain this is
$\tilde{\phi}_{0,6}$ ($\phi_{\lambda=6}$ at the rational points
$\tilde{r}=\sqrt{p/2}$ for $p>6$).  This operator has scaling dimension
$X_{\lambda=6}= h_{\lambda=6}+\bar{h}_{\lambda=6} =36/(2p)$, hence marginal
for $p=9$.  The $p=9$ rational orbifold CFT appears at coupling constant
$\gamma=0$ or $J^z=|J|$ in our parametrization of the anyon chain --- in the
related spin-$1/2$ XXZ Heisenberg chain this is the location of the transition
between the antiferromagnetic critical (disordered) and massive regimes.

Finally we note that, by duality of the orbifold theory, the full sequence of
rational CFTs with any integer $p$ appears at compactification radii
$1/\sqrt{2p}$ or coupling constants
\begin{equation}
  \label{cft-orb2}
  \bar{\gamma}(p) = \pi\left(1-\frac{1}{9p}\right)\,,\quad 
  p=1,2,3,\ldots\,,
\end{equation}
see Fig.~\ref{fig:d3-phases}.  These points accumulate near $\gamma=\pi$,
where the first order transition into the gapped phase with $\mathbb{Z}_3$
sublattice structure takes place.  The identification of the orbifold CFTs at
these couplings by numerical studies, however, is difficult as a consequence
of the diverging density of low energy states in this regime.

\begin{acknowledgments}
  Funding for this project has been provided by the Deutsche
  Forschungsgemeinschaft under Grant No.\ Fr~737/7-1.
\end{acknowledgments}



\begin{thebibliography}{40}%
\makeatletter
\providecommand \@ifxundefined [1]{%
 \@ifx{#1\undefined}
}%
\providecommand \@ifnum [1]{%
 \ifnum #1\expandafter \@firstoftwo
 \else \expandafter \@secondoftwo
 \fi
}%
\providecommand \@ifx [1]{%
 \ifx #1\expandafter \@firstoftwo
 \else \expandafter \@secondoftwo
 \fi
}%
\providecommand \natexlab [1]{#1}%
\providecommand \enquote  [1]{``#1''}%
\providecommand \bibnamefont  [1]{#1}%
\providecommand \bibfnamefont [1]{#1}%
\providecommand \citenamefont [1]{#1}%
\providecommand \href@noop [0]{\@secondoftwo}%
\providecommand \href [0]{\begingroup \@sanitize@url \@href}%
\providecommand \@href[1]{\@@startlink{#1}\@@href}%
\providecommand \@@href[1]{\endgroup#1\@@endlink}%
\providecommand \@sanitize@url [0]{\catcode `\\12\catcode `\$12\catcode
  `\&12\catcode `\#12\catcode `\^12\catcode `\_12\catcode `\%12\relax}%
\providecommand \@@startlink[1]{}%
\providecommand \@@endlink[0]{}%
\providecommand \url  [0]{\begingroup\@sanitize@url \@url }%
\providecommand \@url [1]{\endgroup\@href {#1}{\urlprefix }}%
\providecommand \urlprefix  [0]{URL }%
\providecommand \Eprint [0]{\href }%
\providecommand \doibase [0]{http://dx.doi.org/}%
\providecommand \selectlanguage [0]{\@gobble}%
\providecommand \bibinfo  [0]{\@secondoftwo}%
\providecommand \bibfield  [0]{\@secondoftwo}%
\providecommand \translation [1]{[#1]}%
\providecommand \BibitemOpen [0]{}%
\providecommand \bibitemStop [0]{}%
\providecommand \bibitemNoStop [0]{.\EOS\space}%
\providecommand \EOS [0]{\spacefactor3000\relax}%
\providecommand \BibitemShut  [1]{\csname bibitem#1\endcsname}%
\let\auto@bib@innerbib\@empty
\bibitem [{\citenamefont {Moore}\ and\ \citenamefont {Read}(1991)}]{MoRe91}%
  \BibitemOpen
  \bibfield  {author} {\bibinfo {author} {\bibfnamefont {Gregory}\ \bibnamefont
  {Moore}}\ and\ \bibinfo {author} {\bibfnamefont {Nicholas}\ \bibnamefont
  {Read}},\ }\bibfield  {title} {\enquote {\bibinfo {title} {{N}onabelions in
  the fractional quantum {H}all effect},}\ }\href@noop {} {\bibfield  {journal}
  {\bibinfo  {journal} {Nucl. Phys. B}\ }\textbf {\bibinfo {volume} {360}},\
  \bibinfo {pages} {362--396} (\bibinfo {year} {1991})}\BibitemShut {NoStop}%
\bibitem [{\citenamefont {Moessner}\ and\ \citenamefont
  {Sondhi}(2001)}]{MoSo01}%
  \BibitemOpen
  \bibfield  {author} {\bibinfo {author} {\bibfnamefont {R.}~\bibnamefont
  {Moessner}}\ and\ \bibinfo {author} {\bibfnamefont {S.~L.}\ \bibnamefont
  {Sondhi}},\ }\bibfield  {title} {\enquote {\bibinfo {title} {{A}n {RVB} phase
  in the triangular lattice quantum dimer model},}\ }\href@noop {} {\bibfield
  {journal} {\bibinfo  {journal} {Phys. Rev. Lett.}\ }\textbf {\bibinfo
  {volume} {86}},\ \bibinfo {pages} {1881--1884} (\bibinfo {year} {2001})},\
  \Eprint {http://arxiv.org/abs/cond-mat/0007378} {cond-mat/0007378}
  \BibitemShut {NoStop}%
\bibitem [{\citenamefont {Balents}\ \emph {et~al.}(2002)\citenamefont
  {Balents}, \citenamefont {Fisher},\ and\ \citenamefont {Girvin}}]{BaFG02}%
  \BibitemOpen
  \bibfield  {author} {\bibinfo {author} {\bibfnamefont {L.}~\bibnamefont
  {Balents}}, \bibinfo {author} {\bibfnamefont {M.~P.~A.}\ \bibnamefont
  {Fisher}}, \ and\ \bibinfo {author} {\bibfnamefont {S.~M.}\ \bibnamefont
  {Girvin}},\ }\bibfield  {title} {\enquote {\bibinfo {title}
  {{F}ractionalization in an {E}asy-axis {K}agome {A}ntiferromagnet},}\
  }\href@noop {} {\bibfield  {journal} {\bibinfo  {journal} {Phys. Rev. B}\
  }\textbf {\bibinfo {volume} {65}},\ \bibinfo {pages} {224412} (\bibinfo
  {year} {2002})},\ \Eprint {http://arxiv.org/abs/cond-mat/0110005}
  {cond-mat/0110005} \BibitemShut {NoStop}%
\bibitem [{\citenamefont {Kitaev}(2006)}]{Kita06}%
  \BibitemOpen
  \bibfield  {author} {\bibinfo {author} {\bibfnamefont {Alexei}\ \bibnamefont
  {Kitaev}},\ }\bibfield  {title} {\enquote {\bibinfo {title} {{A}nyons in an
  exactly solved model and beyond},}\ }\href@noop {} {\bibfield  {journal}
  {\bibinfo  {journal} {Ann. Phys. (NY)}\ }\textbf {\bibinfo {volume} {321}},\
  \bibinfo {pages} {2--111} (\bibinfo {year} {2006})},\ \Eprint
  {http://arxiv.org/abs/cond-mat/0506438} {cond-mat/0506438} \BibitemShut
  {NoStop}%
\bibitem [{\citenamefont {Feiguin}\ \emph {et~al.}(2007)\citenamefont
  {Feiguin}, \citenamefont {Trebst}, \citenamefont {Ludwig}, \citenamefont
  {Troyer}, \citenamefont {Kitaev}, \citenamefont {Wang},\ and\ \citenamefont
  {Freedman}}]{FTLT07}%
  \BibitemOpen
  \bibfield  {author} {\bibinfo {author} {\bibfnamefont {Adrian}\ \bibnamefont
  {Feiguin}}, \bibinfo {author} {\bibfnamefont {Simon}\ \bibnamefont {Trebst}},
  \bibinfo {author} {\bibfnamefont {Andreas W.~W.}\ \bibnamefont {Ludwig}},
  \bibinfo {author} {\bibfnamefont {Matthias}\ \bibnamefont {Troyer}}, \bibinfo
  {author} {\bibfnamefont {Alexei}\ \bibnamefont {Kitaev}}, \bibinfo {author}
  {\bibfnamefont {Zhenghan}\ \bibnamefont {Wang}}, \ and\ \bibinfo {author}
  {\bibfnamefont {Michael~H.}\ \bibnamefont {Freedman}},\ }\bibfield  {title}
  {\enquote {\bibinfo {title} {{I}nteracting anyons in topological quantum
  liquids: {T}he golden chain},}\ }\href@noop {} {\bibfield  {journal}
  {\bibinfo  {journal} {Phys. Rev. Lett.}\ }\textbf {\bibinfo {volume} {98}},\
  \bibinfo {pages} {160409} (\bibinfo {year} {2007})},\ \Eprint
  {http://arxiv.org/abs/cond-mat/0612341} {cond-mat/0612341} \BibitemShut
  {NoStop}%
\bibitem [{\citenamefont {Trebst}\ \emph {et~al.}(2008)\citenamefont {Trebst},
  \citenamefont {Ardonne}, \citenamefont {Feiguin}, \citenamefont {Huse},
  \citenamefont {Ludwig},\ and\ \citenamefont {Troyer}}]{TAFH08}%
  \BibitemOpen
  \bibfield  {author} {\bibinfo {author} {\bibfnamefont {Simon}\ \bibnamefont
  {Trebst}}, \bibinfo {author} {\bibfnamefont {Eddy}\ \bibnamefont {Ardonne}},
  \bibinfo {author} {\bibfnamefont {Adrian}\ \bibnamefont {Feiguin}}, \bibinfo
  {author} {\bibfnamefont {David~A.}\ \bibnamefont {Huse}}, \bibinfo {author}
  {\bibfnamefont {Andreas W.~W.}\ \bibnamefont {Ludwig}}, \ and\ \bibinfo
  {author} {\bibfnamefont {Matthias}\ \bibnamefont {Troyer}},\ }\bibfield
  {title} {\enquote {\bibinfo {title} {{C}ollective states of interacting
  {F}ibonacci anyons},}\ }\href@noop {} {\bibfield  {journal} {\bibinfo
  {journal} {Phys. Rev. Lett.}\ }\textbf {\bibinfo {volume} {101}},\ \bibinfo
  {pages} {050401} (\bibinfo {year} {2008})},\ \Eprint
  {http://arxiv.org/abs/0801.4602} {arXiv:0801.4602} \BibitemShut {NoStop}%
\bibitem [{\citenamefont {Gils}\ \emph {et~al.}(2013)\citenamefont {Gils},
  \citenamefont {Ardonne}, \citenamefont {Trebst}, \citenamefont {Huse},
  \citenamefont {Ludwig}, \citenamefont {Troyer},\ and\ \citenamefont
  {Wang}}]{GATH13}%
  \BibitemOpen
  \bibfield  {author} {\bibinfo {author} {\bibfnamefont {Charlotte}\
  \bibnamefont {Gils}}, \bibinfo {author} {\bibfnamefont {Eddy}\ \bibnamefont
  {Ardonne}}, \bibinfo {author} {\bibfnamefont {Simon}\ \bibnamefont {Trebst}},
  \bibinfo {author} {\bibfnamefont {David~A.}\ \bibnamefont {Huse}}, \bibinfo
  {author} {\bibfnamefont {Andreas W.~W.}\ \bibnamefont {Ludwig}}, \bibinfo
  {author} {\bibfnamefont {Matthias}\ \bibnamefont {Troyer}}, \ and\ \bibinfo
  {author} {\bibfnamefont {Zhenghan}\ \bibnamefont {Wang}},\ }\bibfield
  {title} {\enquote {\bibinfo {title} {{A}nyonic quantum spin chains: {S}pin-1
  generalizations and topological stability},}\ }\href@noop {} {\bibfield
  {journal} {\bibinfo  {journal} {Phys. Rev. B}\ }\textbf {\bibinfo {volume}
  {87}},\ \bibinfo {pages} {235120} (\bibinfo {year} {2013})},\ \Eprint
  {http://arxiv.org/abs/1303.4290} {arXiv:1303.4290} \BibitemShut {NoStop}%
\bibitem [{\citenamefont {Finch}\ \emph
  {et~al.}(2014{\natexlab{a}})\citenamefont {Finch}, \citenamefont {Frahm},
  \citenamefont {Lewerenz}, \citenamefont {Milsted},\ and\ \citenamefont
  {Osborne}}]{Finch.etal14}%
  \BibitemOpen
  \bibfield  {author} {\bibinfo {author} {\bibfnamefont {Peter~E.}\
  \bibnamefont {Finch}}, \bibinfo {author} {\bibfnamefont {Holger}\
  \bibnamefont {Frahm}}, \bibinfo {author} {\bibfnamefont {Marius}\
  \bibnamefont {Lewerenz}}, \bibinfo {author} {\bibfnamefont {Ashley}\
  \bibnamefont {Milsted}}, \ and\ \bibinfo {author} {\bibfnamefont {Tobias~J.}\
  \bibnamefont {Osborne}},\ }\bibfield  {title} {\enquote {\bibinfo {title}
  {{Q}uantum phases of a chain of strongly interacting anyons},}\ }\href@noop
  {} {\bibfield  {journal} {\bibinfo  {journal} {Phys. Rev. B}\ }\textbf
  {\bibinfo {volume} {90}},\ \bibinfo {pages} {081111(R)} (\bibinfo {year}
  {2014}{\natexlab{a}})},\ \Eprint {http://arxiv.org/abs/1404.2439}
  {arXiv:1404.2439} \BibitemShut {NoStop}%
\bibitem [{\citenamefont {Gils}\ \emph {et~al.}(2009)\citenamefont {Gils},
  \citenamefont {Ardonne}, \citenamefont {Trebst}, \citenamefont {Ludwig},
  \citenamefont {Troyer},\ and\ \citenamefont {Wang}}]{GATL09}%
  \BibitemOpen
  \bibfield  {author} {\bibinfo {author} {\bibfnamefont {Charlotte}\
  \bibnamefont {Gils}}, \bibinfo {author} {\bibfnamefont {Eddy}\ \bibnamefont
  {Ardonne}}, \bibinfo {author} {\bibfnamefont {Simon}\ \bibnamefont {Trebst}},
  \bibinfo {author} {\bibfnamefont {Andreas W.~W.}\ \bibnamefont {Ludwig}},
  \bibinfo {author} {\bibfnamefont {Matthias}\ \bibnamefont {Troyer}}, \ and\
  \bibinfo {author} {\bibfnamefont {Zhenghan}\ \bibnamefont {Wang}},\
  }\bibfield  {title} {\enquote {\bibinfo {title} {{C}ollective {S}tates of
  {I}nteracting {A}nyons, {E}dge {S}tates, and the {N}ucleation of
  {T}opological {L}iquids},}\ }\href@noop {} {\bibfield  {journal} {\bibinfo
  {journal} {Phys. Rev. Lett.}\ }\textbf {\bibinfo {volume} {103}},\ \bibinfo
  {pages} {070401} (\bibinfo {year} {2009})},\ \Eprint
  {http://arxiv.org/abs/0810.2277} {arXiv:0810.2277} \BibitemShut {NoStop}%
\bibitem [{\citenamefont {Grosfeld}\ and\ \citenamefont
  {Schoutens}(2009)}]{GrSc09}%
  \BibitemOpen
  \bibfield  {author} {\bibinfo {author} {\bibfnamefont {Eytan}\ \bibnamefont
  {Grosfeld}}\ and\ \bibinfo {author} {\bibfnamefont {Kareljan}\ \bibnamefont
  {Schoutens}},\ }\bibfield  {title} {\enquote {\bibinfo {title}
  {{N}on-{A}belian anyons: when {I}sing meets {F}ibonacci},}\ }\href@noop {}
  {\bibfield  {journal} {\bibinfo  {journal} {Phys. Rev. Lett.}\ }\textbf
  {\bibinfo {volume} {103}},\ \bibinfo {pages} {076803} (\bibinfo {year}
  {2009})},\ \Eprint {http://arxiv.org/abs/0810.1955} {arXiv:0810.1955}
  \BibitemShut {NoStop}%
\bibitem [{\citenamefont {Bais}\ and\ \citenamefont
  {Slingerland}(2009)}]{BaSl09}%
  \BibitemOpen
  \bibfield  {author} {\bibinfo {author} {\bibfnamefont {F.~A.}\ \bibnamefont
  {Bais}}\ and\ \bibinfo {author} {\bibfnamefont {J.~K.}\ \bibnamefont
  {Slingerland}},\ }\bibfield  {title} {\enquote {\bibinfo {title}
  {{C}ondensate-induced transitions between topologically ordered phases},}\
  }\href@noop {} {\bibfield  {journal} {\bibinfo  {journal} {Phys. Rev. B}\
  }\textbf {\bibinfo {volume} {79}},\ \bibinfo {pages} {045316} (\bibinfo
  {year} {2009})},\ \Eprint {http://arxiv.org/abs/0808.0627} {arXiv:0808.0627}
  \BibitemShut {NoStop}%
\bibitem [{\citenamefont {Andrews}\ \emph {et~al.}(1984)\citenamefont
  {Andrews}, \citenamefont {Baxter},\ and\ \citenamefont {Forrester}}]{AnBF84}%
  \BibitemOpen
  \bibfield  {author} {\bibinfo {author} {\bibfnamefont {G.~E.}\ \bibnamefont
  {Andrews}}, \bibinfo {author} {\bibfnamefont {R.~J.}\ \bibnamefont {Baxter}},
  \ and\ \bibinfo {author} {\bibfnamefont {P.~J.}\ \bibnamefont {Forrester}},\
  }\bibfield  {title} {\enquote {\bibinfo {title} {{E}ight-vertex {SOS} model
  and generalized {R}ogers-{R}amanujan-type identities},}\ }\href@noop {}
  {\bibfield  {journal} {\bibinfo  {journal} {J. Stat. Phys.}\ }\textbf
  {\bibinfo {volume} {35}},\ \bibinfo {pages} {193--266} (\bibinfo {year}
  {1984})}\BibitemShut {NoStop}%
\bibitem [{\citenamefont {Pasquier}(1988)}]{Pasq88}%
  \BibitemOpen
  \bibfield  {author} {\bibinfo {author} {\bibfnamefont {V.}~\bibnamefont
  {Pasquier}},\ }\bibfield  {title} {\enquote {\bibinfo {title} {{E}tiology of
  {IRF} models},}\ }\href@noop {} {\bibfield  {journal} {\bibinfo  {journal}
  {Comm. Math. Phys.}\ }\textbf {\bibinfo {volume} {118}},\ \bibinfo {pages}
  {355--364} (\bibinfo {year} {1988})}\BibitemShut {NoStop}%
\bibitem [{\citenamefont {Gepner}(1992)}]{Gepn92}%
  \BibitemOpen
  \bibfield  {author} {\bibinfo {author} {\bibfnamefont {Doron}\ \bibnamefont
  {Gepner}},\ }\href@noop {} {\emph {\bibinfo {title} {{F}oundations of
  {R}ational {Q}uantum {F}ield {T}heory, {I}}}},\ \bibinfo {type} {Preprint}\
  \bibinfo {number} {CALT-68-1825}\ (\bibinfo  {institution} {Caltech},\
  \bibinfo {year} {1992})\ \Eprint {http://arxiv.org/abs/hep-th/9211100}
  {hep-th/9211100} \BibitemShut {NoStop}%
\bibitem [{\citenamefont {Finch}\ and\ \citenamefont {Frahm}(2013)}]{FiFr13}%
  \BibitemOpen
  \bibfield  {author} {\bibinfo {author} {\bibfnamefont {Peter~E.}\
  \bibnamefont {Finch}}\ and\ \bibinfo {author} {\bibfnamefont {Holger}\
  \bibnamefont {Frahm}},\ }\bibfield  {title} {\enquote {\bibinfo {title}
  {{T}he ${D}({D}_{3})$-anyon chain: integrable boundary conditions and
  excitation spectra},}\ }\href@noop {} {\bibfield  {journal} {\bibinfo
  {journal} {New J. Phys.}\ }\textbf {\bibinfo {volume} {15}},\ \bibinfo
  {pages} {053035} (\bibinfo {year} {2013})},\ \Eprint
  {http://arxiv.org/abs/1211.4449} {arXiv:1211.4449} \BibitemShut {NoStop}%
\bibitem [{\citenamefont {Finch}\ \emph
  {et~al.}(2014{\natexlab{b}})\citenamefont {Finch}, \citenamefont {Flohr},\
  and\ \citenamefont {Frahm}}]{FiFF14}%
  \BibitemOpen
  \bibfield  {author} {\bibinfo {author} {\bibfnamefont {Peter~E.}\
  \bibnamefont {Finch}}, \bibinfo {author} {\bibfnamefont {Michael}\
  \bibnamefont {Flohr}}, \ and\ \bibinfo {author} {\bibfnamefont {Holger}\
  \bibnamefont {Frahm}},\ }\bibfield  {title} {\enquote {\bibinfo {title}
  {{I}ntegrable anyon chains: from fusion rules to face models to effective
  field theories},}\ }\href@noop {} {\bibfield  {journal} {\bibinfo  {journal}
  {Nucl. Phys. B}\ }\textbf {\bibinfo {volume} {889}},\ \bibinfo {pages}
  {299--332} (\bibinfo {year} {2014}{\natexlab{b}})},\ \Eprint
  {http://arxiv.org/abs/1408.1282} {arXiv:1408.1282} \BibitemShut {NoStop}%
\bibitem [{\citenamefont {Martina}\ \emph {et~al.}(2010)\citenamefont
  {Martina}, \citenamefont {Protogenov},\ and\ \citenamefont
  {Verbus}}]{MaPV10}%
  \BibitemOpen
  \bibfield  {author} {\bibinfo {author} {\bibfnamefont {L.}~\bibnamefont
  {Martina}}, \bibinfo {author} {\bibfnamefont {A.}~\bibnamefont {Protogenov}},
  \ and\ \bibinfo {author} {\bibfnamefont {V.}~\bibnamefont {Verbus}},\
  }\bibfield  {title} {\enquote {\bibinfo {title} {{A} chain of strongly
  correlated ${SU}(2)_4$ anyons: {H}amiltonian and {H}ilbert space of
  states},}\ }\href@noop {} {\bibfield  {journal} {\bibinfo  {journal}
  {preprint}\ } (\bibinfo {year} {2010})},\ \Eprint
  {http://arxiv.org/abs/1001.4932} {arXiv:1001.4932} \BibitemShut {NoStop}%
\bibitem [{\citenamefont {Verbus}\ \emph {et~al.}(2011)\citenamefont {Verbus},
  \citenamefont {Martina},\ and\ \citenamefont {Protogenov}}]{VeMP11}%
  \BibitemOpen
  \bibfield  {author} {\bibinfo {author} {\bibfnamefont {V.~A.}\ \bibnamefont
  {Verbus}}, \bibinfo {author} {\bibfnamefont {L.}~\bibnamefont {Martina}}, \
  and\ \bibinfo {author} {\bibfnamefont {A.~P.}\ \bibnamefont {Protogenov}},\
  }\bibfield  {title} {\enquote {\bibinfo {title} {{C}hain of interacting
  ${SU}(2)_4$ anyons and quantum ${SU}(2)_k \times\overline{SU(2)_k}$
  doubles},}\ }\href@noop {} {\bibfield  {journal} {\bibinfo  {journal} {Theor.
  Math. Phys.}\ }\textbf {\bibinfo {volume} {167}},\ \bibinfo {pages}
  {843--855} (\bibinfo {year} {2011})}\BibitemShut {NoStop}%
\bibitem [{\citenamefont {Finch}(2013)}]{Finch13}%
  \BibitemOpen
  \bibfield  {author} {\bibinfo {author} {\bibfnamefont {Peter~E.}\
  \bibnamefont {Finch}},\ }\bibfield  {title} {\enquote {\bibinfo {title}
  {{F}rom spin to anyon notation: {T}he {XXZ} {H}eisenberg model as a ${D}_{3}$
  (or $su(2)_{4}$) anyon chain},}\ }\href@noop {} {\bibfield  {journal}
  {\bibinfo  {journal} {J. Phys. A}\ }\textbf {\bibinfo {volume} {46}},\
  \bibinfo {pages} {055305} (\bibinfo {year} {2013})},\ \Eprint
  {http://arxiv.org/abs/1201.4470} {arXiv:1201.4470} \BibitemShut {NoStop}%
\bibitem [{\citenamefont {Roche}(1990)}]{Roch90}%
  \BibitemOpen
  \bibfield  {author} {\bibinfo {author} {\bibfnamefont {{\relax
  Ph}.}~\bibnamefont {Roche}},\ }\bibfield  {title} {\enquote {\bibinfo {title}
  {{O}cneanu cell calculus and integrable lattice models},}\ }\href@noop {}
  {\bibfield  {journal} {\bibinfo  {journal} {Comm. Math. Phys.}\ }\textbf
  {\bibinfo {volume} {127}},\ \bibinfo {pages} {395--424} (\bibinfo {year}
  {1990})}\BibitemShut {NoStop}%
\bibitem [{\citenamefont {Kohmoto}\ \emph {et~al.}(1981)\citenamefont
  {Kohmoto}, \citenamefont {den Nijs},\ and\ \citenamefont
  {Kadanoff}}]{KoNK81}%
  \BibitemOpen
  \bibfield  {author} {\bibinfo {author} {\bibfnamefont {Mahito}\ \bibnamefont
  {Kohmoto}}, \bibinfo {author} {\bibfnamefont {Marcel}\ \bibnamefont {den
  Nijs}}, \ and\ \bibinfo {author} {\bibfnamefont {Leo~P.}\ \bibnamefont
  {Kadanoff}},\ }\bibfield  {title} {\enquote {\bibinfo {title} {{H}amiltonian
  studies of the $d=2$ {Ashkin-Teller} model},}\ }\href@noop {} {\bibfield
  {journal} {\bibinfo  {journal} {Phys. Rev. B}\ }\textbf {\bibinfo {volume}
  {24}},\ \bibinfo {pages} {5229--5241} (\bibinfo {year} {1981})}\BibitemShut
  {NoStop}%
\bibitem [{\citenamefont {Yang}(1987)}]{Yang87}%
  \BibitemOpen
  \bibfield  {author} {\bibinfo {author} {\bibfnamefont {Sung-Kil}\
  \bibnamefont {Yang}},\ }\bibfield  {title} {\enquote {\bibinfo {title}
  {{M}odular invariant partition function of the {A}shkin-{T}eller model on the
  critical line and ${N} = 2$ superconformal invariance},}\ }\href@noop {}
  {\bibfield  {journal} {\bibinfo  {journal} {Nucl. Phys. B}\ }\textbf
  {\bibinfo {volume} {285}},\ \bibinfo {pages} {183--203} (\bibinfo {year}
  {1987})}\BibitemShut {NoStop}%
\bibitem [{\citenamefont {Saleur}(1987)}]{Saleur_AT87}%
  \BibitemOpen
  \bibfield  {author} {\bibinfo {author} {\bibfnamefont {H.}~\bibnamefont
  {Saleur}},\ }\bibfield  {title} {\enquote {\bibinfo {title} {{P}artition
  functions of the two-dimensional {A}shkin-{T}eller model on the critical
  line},}\ }\href@noop {} {\bibfield  {journal} {\bibinfo  {journal} {J. Phys.
  A}\ }\textbf {\bibinfo {volume} {20}},\ \bibinfo {pages} {L1127--L1133}
  (\bibinfo {year} {1987})}\BibitemShut {NoStop}%
\bibitem [{\citenamefont {di~Francesco}\ and\ \citenamefont
  {Zuber}(1990)}]{FrZu90}%
  \BibitemOpen
  \bibfield  {author} {\bibinfo {author} {\bibfnamefont {P.}~\bibnamefont
  {di~Francesco}}\ and\ \bibinfo {author} {\bibfnamefont {J.-B.}\ \bibnamefont
  {Zuber}},\ }\bibfield  {title} {\enquote {\bibinfo {title} {{SU}({N}) lattice
  integrable models associated with graphs},}\ }\href@noop {} {\bibfield
  {journal} {\bibinfo  {journal} {Nucl. Phys. B}\ }\textbf {\bibinfo {volume}
  {338}},\ \bibinfo {pages} {602--646} (\bibinfo {year} {1990})}\BibitemShut
  {NoStop}%
\bibitem [{\citenamefont {Kl{\"u}mper}\ and\ \citenamefont
  {Pearce}(1992)}]{KlPe92}%
  \BibitemOpen
  \bibfield  {author} {\bibinfo {author} {\bibfnamefont {Andreas}\ \bibnamefont
  {Kl{\"u}mper}}\ and\ \bibinfo {author} {\bibfnamefont {Paul~A.}\ \bibnamefont
  {Pearce}},\ }\bibfield  {title} {\enquote {\bibinfo {title} {{C}onformal
  weights of {RSOS} lattice models and their fusion hierarchies},}\ }\href@noop
  {} {\bibfield  {journal} {\bibinfo  {journal} {Physica A}\ }\textbf {\bibinfo
  {volume} {183}},\ \bibinfo {pages} {304--350} (\bibinfo {year}
  {1992})}\BibitemShut {NoStop}%
\bibitem [{\citenamefont {Frahm}\ and\ \citenamefont
  {Karaiskos}(2014)}]{FrKa14}%
  \BibitemOpen
  \bibfield  {author} {\bibinfo {author} {\bibfnamefont {Holger}\ \bibnamefont
  {Frahm}}\ and\ \bibinfo {author} {\bibfnamefont {Nikos}\ \bibnamefont
  {Karaiskos}},\ }\bibfield  {title} {\enquote {\bibinfo {title} {{I}nversion
  identities for inhomogeneous face models},}\ }\href@noop {} {\bibfield
  {journal} {\bibinfo  {journal} {Nucl. Phys. B}\ }\textbf {\bibinfo {volume}
  {887}},\ \bibinfo {pages} {423--440} (\bibinfo {year} {2014})},\ \Eprint
  {http://arxiv.org/abs/1407.6883} {arXiv:1407.6883} \BibitemShut {NoStop}%
\bibitem [{\citenamefont {Frahm}\ and\ \citenamefont
  {Karaiskos}(2015)}]{FrKa15}%
  \BibitemOpen
  \bibfield  {author} {\bibinfo {author} {\bibfnamefont {Holger}\ \bibnamefont
  {Frahm}}\ and\ \bibinfo {author} {\bibfnamefont {Nikos}\ \bibnamefont
  {Karaiskos}},\ }\bibfield  {title} {\enquote {\bibinfo {title}
  {{N}on-{A}belian ${SU}(3)_k$ anyons: inversion identities for higher rank
  face models},}\ }\href@noop {} {\bibfield  {journal} {\bibinfo  {journal} {J.
  Phys. A}\ }\textbf {\bibinfo {volume} {48}},\ \bibinfo {pages} {484001}
  (\bibinfo {year} {2015})},\ \Eprint {http://arxiv.org/abs/1506.00822}
  {arXiv:1506.00822} \BibitemShut {NoStop}%
\bibitem [{\citenamefont {Baxter}(1982)}]{Baxter:book}%
  \BibitemOpen
  \bibfield  {author} {\bibinfo {author} {\bibfnamefont {R.~J.}\ \bibnamefont
  {Baxter}},\ }\href@noop {} {\emph {\bibinfo {title} {{E}xactly {S}olved
  {M}odels in {S}tatistical {M}echanics}}}\ (\bibinfo  {publisher} {Academic
  Press},\ \bibinfo {address} {London},\ \bibinfo {year} {1982})\BibitemShut
  {NoStop}%
\bibitem [{\citenamefont {Batchelor}\ \emph {et~al.}(1995)\citenamefont
  {Batchelor}, \citenamefont {Baxter}, \citenamefont {O'Rourke},\ and\
  \citenamefont {Yung}}]{BBOY95}%
  \BibitemOpen
  \bibfield  {author} {\bibinfo {author} {\bibfnamefont {M.~T.}\ \bibnamefont
  {Batchelor}}, \bibinfo {author} {\bibfnamefont {R.~J.}\ \bibnamefont
  {Baxter}}, \bibinfo {author} {\bibfnamefont {M.~J.}\ \bibnamefont
  {O'Rourke}}, \ and\ \bibinfo {author} {\bibfnamefont {C.~M.}\ \bibnamefont
  {Yung}},\ }\bibfield  {title} {\enquote {\bibinfo {title} {{E}xact solution
  and interfacial tension of the six-vertex model with anti-periodic boundary
  conditions},}\ }\href@noop {} {\bibfield  {journal} {\bibinfo  {journal} {J.
  Phys. A}\ }\textbf {\bibinfo {volume} {28}},\ \bibinfo {pages} {2759--2770}
  (\bibinfo {year} {1995})},\ \Eprint {http://arxiv.org/abs/hep-th/9502040}
  {hep-th/9502040} \BibitemShut {NoStop}%
\bibitem [{\citenamefont {Niekamp}\ \emph {et~al.}(2009)\citenamefont
  {Niekamp}, \citenamefont {Wirth},\ and\ \citenamefont {Frahm}}]{NiWF09}%
  \BibitemOpen
  \bibfield  {author} {\bibinfo {author} {\bibfnamefont {S{\"o}nke}\
  \bibnamefont {Niekamp}}, \bibinfo {author} {\bibfnamefont {Tobias}\
  \bibnamefont {Wirth}}, \ and\ \bibinfo {author} {\bibfnamefont {Holger}\
  \bibnamefont {Frahm}},\ }\bibfield  {title} {\enquote {\bibinfo {title}
  {{T}he {XXZ} model with anti-periodic twisted boundary conditions},}\
  }\href@noop {} {\bibfield  {journal} {\bibinfo  {journal} {J. Phys. A: Math.
  Theor.}\ }\textbf {\bibinfo {volume} {42}},\ \bibinfo {pages} {195008}
  (\bibinfo {year} {2009})},\ \Eprint {http://arxiv.org/abs/0902.1079}
  {arXiv:0902.1079} \BibitemShut {NoStop}%
\bibitem [{\citenamefont {Yang}\ and\ \citenamefont {Yang}(1966)}]{YaYa66b}%
  \BibitemOpen
  \bibfield  {author} {\bibinfo {author} {\bibfnamefont {C.~N.}\ \bibnamefont
  {Yang}}\ and\ \bibinfo {author} {\bibfnamefont {C.~P.}\ \bibnamefont
  {Yang}},\ }\bibfield  {title} {\enquote {\bibinfo {title} {{O}ne-dimensional
  chain of anisotropic spin-spin interactions. {II}. {P}roperties of the
  ground-state energy per lattice site for an infinite system},}\ }\href@noop
  {} {\bibfield  {journal} {\bibinfo  {journal} {Phys. Rev.}\ }\textbf
  {\bibinfo {volume} {150}},\ \bibinfo {pages} {327--339} (\bibinfo {year}
  {1966})}\BibitemShut {NoStop}%
\bibitem [{\citenamefont {Avdeev}\ and\ \citenamefont
  {D\"orfel}(1986)}]{AvDo86}%
  \BibitemOpen
  \bibfield  {author} {\bibinfo {author} {\bibfnamefont {L.~V.}\ \bibnamefont
  {Avdeev}}\ and\ \bibinfo {author} {\bibfnamefont {B.-D.}\ \bibnamefont
  {D\"orfel}},\ }\bibfield  {title} {\enquote {\bibinfo {title} {{F}inite-size
  corrections for the {XXX} antiferromagnet},}\ }\href@noop {} {\bibfield
  {journal} {\bibinfo  {journal} {J. Phys. A}\ }\textbf {\bibinfo {volume}
  {19}},\ \bibinfo {pages} {L13--L17} (\bibinfo {year} {1986})}\BibitemShut
  {NoStop}%
\bibitem [{\citenamefont {Hamer}(1986)}]{Hame86}%
  \BibitemOpen
  \bibfield  {author} {\bibinfo {author} {\bibfnamefont {C.~J.}\ \bibnamefont
  {Hamer}},\ }\bibfield  {title} {\enquote {\bibinfo {title} {{F}inite-size
  corrections for ground states of the {XXZ} {H}eisenberg chain},}\ }\href@noop
  {} {\bibfield  {journal} {\bibinfo  {journal} {J. Phys. A}\ }\textbf
  {\bibinfo {volume} {19}},\ \bibinfo {pages} {3335--3351} (\bibinfo {year}
  {1986})}\BibitemShut {NoStop}%
\bibitem [{\citenamefont {Bl{\"o}te}\ \emph {et~al.}(1986)\citenamefont
  {Bl{\"o}te}, \citenamefont {Cardy},\ and\ \citenamefont
  {Nightingale}}]{BlCN86}%
  \BibitemOpen
  \bibfield  {author} {\bibinfo {author} {\bibfnamefont {H.~W.~J.}\
  \bibnamefont {Bl{\"o}te}}, \bibinfo {author} {\bibfnamefont {John~L.}\
  \bibnamefont {Cardy}}, \ and\ \bibinfo {author} {\bibfnamefont {M.~P.}\
  \bibnamefont {Nightingale}},\ }\bibfield  {title} {\enquote {\bibinfo {title}
  {{C}onformal invariance, the central charge and universal finite-size
  amplitudes at criticality},}\ }\href@noop {} {\bibfield  {journal} {\bibinfo
  {journal} {Phys. Rev. Lett.}\ }\textbf {\bibinfo {volume} {56}},\ \bibinfo
  {pages} {742--745} (\bibinfo {year} {1986})}\BibitemShut {NoStop}%
\bibitem [{\citenamefont {Affleck}(1986)}]{Affl86}%
  \BibitemOpen
  \bibfield  {author} {\bibinfo {author} {\bibfnamefont {Ian}\ \bibnamefont
  {Affleck}},\ }\bibfield  {title} {\enquote {\bibinfo {title} {{U}niversal
  term in the free energy at a critical point and and the conformal anomaly},}\
  }\href@noop {} {\bibfield  {journal} {\bibinfo  {journal} {Phys. Rev. Lett.}\
  }\textbf {\bibinfo {volume} {56}},\ \bibinfo {pages} {746--748} (\bibinfo
  {year} {1986})}\BibitemShut {NoStop}%
\bibitem [{\citenamefont {Cardy}(1986)}]{Card86a}%
  \BibitemOpen
  \bibfield  {author} {\bibinfo {author} {\bibfnamefont {John~L.}\ \bibnamefont
  {Cardy}},\ }\bibfield  {title} {\enquote {\bibinfo {title} {{O}perator
  content of two-dimensional conformally invariant theories},}\ }\href@noop {}
  {\bibfield  {journal} {\bibinfo  {journal} {Nucl. Phys. B}\ }\textbf
  {\bibinfo {volume} {270}},\ \bibinfo {pages} {186--204} (\bibinfo {year}
  {1986})}\BibitemShut {NoStop}%
\bibitem [{\citenamefont {Alcaraz}\ \emph
  {et~al.}(1988{\natexlab{a}})\citenamefont {Alcaraz}, \citenamefont {Barber},\
  and\ \citenamefont {Batchelor}}]{AlBB88}%
  \BibitemOpen
  \bibfield  {author} {\bibinfo {author} {\bibfnamefont {Francisco~C.}\
  \bibnamefont {Alcaraz}}, \bibinfo {author} {\bibfnamefont {Michael~N.}\
  \bibnamefont {Barber}}, \ and\ \bibinfo {author} {\bibfnamefont {Murray~T.}\
  \bibnamefont {Batchelor}},\ }\bibfield  {title} {\enquote {\bibinfo {title}
  {{C}onformal invariance, the {XXZ} chain and the operator content of
  two-dimensional critical systems},}\ }\href@noop {} {\bibfield  {journal}
  {\bibinfo  {journal} {Ann. Phys. (NY)}\ }\textbf {\bibinfo {volume} {182}},\
  \bibinfo {pages} {280--343} (\bibinfo {year}
  {1988}{\natexlab{a}})}\BibitemShut {NoStop}%
\bibitem [{\citenamefont {Alcaraz}\ \emph
  {et~al.}(1988{\natexlab{b}})\citenamefont {Alcaraz}, \citenamefont {Baake},
  \citenamefont {Grimmn},\ and\ \citenamefont {Rittenberg}}]{ABGR88}%
  \BibitemOpen
  \bibfield  {author} {\bibinfo {author} {\bibfnamefont {F.~C.}\ \bibnamefont
  {Alcaraz}}, \bibinfo {author} {\bibfnamefont {M.}~\bibnamefont {Baake}},
  \bibinfo {author} {\bibfnamefont {U.}~\bibnamefont {Grimmn}}, \ and\ \bibinfo
  {author} {\bibfnamefont {V.}~\bibnamefont {Rittenberg}},\ }\bibfield  {title}
  {\enquote {\bibinfo {title} {{O}perator content of the {XXZ} chain},}\
  }\href@noop {} {\bibfield  {journal} {\bibinfo  {journal} {J. Phys. A}\
  }\textbf {\bibinfo {volume} {21}},\ \bibinfo {pages} {L117--L120} (\bibinfo
  {year} {1988}{\natexlab{b}})}\BibitemShut {NoStop}%
\bibitem [{\citenamefont {Ginsparg}(1988)}]{Ginsparg88}%
  \BibitemOpen
  \bibfield  {author} {\bibinfo {author} {\bibfnamefont {Paul~H.}\ \bibnamefont
  {Ginsparg}},\ }\bibfield  {title} {\enquote {\bibinfo {title} {{Curiosities
  at $c = 1$}},}\ }\href {\doibase 10.1016/0550-3213(88)90249-0} {\bibfield
  {journal} {\bibinfo  {journal} {Nucl. Phys. B}\ }\textbf {\bibinfo {volume}
  {295}},\ \bibinfo {pages} {153--170} (\bibinfo {year} {1988})}\BibitemShut
  {NoStop}%
\bibitem [{\citenamefont {Dijkgraaf}\ \emph {et~al.}(1989)\citenamefont
  {Dijkgraaf}, \citenamefont {Vafa}, \citenamefont {Verlinde},\ and\
  \citenamefont {Verlinde}}]{DVVV89}%
  \BibitemOpen
  \bibfield  {author} {\bibinfo {author} {\bibfnamefont {Robbert}\ \bibnamefont
  {Dijkgraaf}}, \bibinfo {author} {\bibfnamefont {Cumrun}\ \bibnamefont
  {Vafa}}, \bibinfo {author} {\bibfnamefont {Erik}\ \bibnamefont {Verlinde}}, \
  and\ \bibinfo {author} {\bibfnamefont {Herman}\ \bibnamefont {Verlinde}},\
  }\bibfield  {title} {\enquote {\bibinfo {title} {{T}he operator algebra of
  orbifold models},}\ }\href@noop {} {\bibfield  {journal} {\bibinfo  {journal}
  {Comm. Math. Phys.}\ }\textbf {\bibinfo {volume} {123}},\ \bibinfo {pages}
  {485--526} (\bibinfo {year} {1989})}\BibitemShut {NoStop}%
\end{thebibliography}
%

\end{document}